\documentclass[twocolumn]{aastex62}

\def\cof {$^{12}\mathrm{CO}~(J=1\rightarrow0)$}
\def\cos {$^{13}\mathrm{CO}~(J=1\rightarrow0)$}
\def\cot {$\mathrm{C}^{18}\mathrm{O}~(J=1\rightarrow0)$}
\def\msun{$M_{\odot}$}
 
\def\ag{$A_{G}$}
\def\cofs {$^{12}\mathrm{CO}$}
\def\coss {$^{13}\mathrm{CO}$}
\def\cots{$\mathrm{C}^{18}\mathrm{O}$}
\def\deg  {\ifmmode {^\circ}\else {$^\circ$}\fi}

\def\kms     {km~s$^{-1}$}
\newcommand{\HII}{\mbox{H\,\textsc{ii}}}%

\newcommand{\para}{$\varpi$}
\newcommand{\dis}{$D$}
\newcommand{\paraerr}{$\Delta \varpi$}

\usepackage{booktabs}
\usepackage{chngcntr}

\date{\today}
\submitjournal{ApJ}

\shorttitle{Molecular cloud distances}
\shortauthors{Yan et al.}

\begin{document}

\title{Molecular cloud distances based on the MWISP CO survey and Gaia DR2} 

\correspondingauthor{Ji Yang}
\email{jiyang@pmo.ac.cn,qzyan@pmo.ac.cn}

\author[0000-0003-4586-7751]{Qing-Zeng Yan}
\affil{Purple Mountain Observatory and Key Laboratory of Radio Astronomy,\\
 Chinese Academy of Sciences, Nanjing 210034, People's Republic of China}

\author{Ji Yang}
\affil{Purple Mountain Observatory and Key Laboratory of Radio Astronomy,\\
 Chinese Academy of Sciences, Nanjing 210034, People's Republic of China}

 \author{Yan Sun}
 \affil{Purple Mountain Observatory and Key Laboratory of Radio Astronomy,\\
  Chinese Academy of Sciences, Nanjing 210034, People's Republic of China}

 \author{Yang Su }
 \affil{Purple Mountain Observatory and Key Laboratory of Radio Astronomy,\\
  Chinese Academy of Sciences, Nanjing 210034, People's Republic of China}
 
 \author{Ye Xu }
 \affil{Purple Mountain Observatory and Key Laboratory of Radio Astronomy,\\
  Chinese Academy of Sciences, Nanjing 210034, People's Republic of China}
  



\begin{abstract}
We present a new method of calculating distances of molecular clouds in the Galactic plane, using CO observations and the \textit{Gaia} DR2 parallax and $G$-band extinction (\ag) measurements. Due to the complexity of dust environments in the Galactic plane, \ag\ contains irregular variations, which is difficult to model. To overcome this difficulty, we propose  that the  \ag\ of off-cloud stars (\textit{Gaia} stars around molecular clouds) can be used   as a baseline to calibrate the \ag\ of on-cloud stars (\textit{Gaia} stars toward molecular clouds), which removes the \ag\ components that are  unrelated to molecular clouds. The distance is subsequently inferred from the jump point in on-cloud \ag\ with Bayesian analysis and Markov chain Monte Carlo (MCMC) sampling.  We applied this baseline subtraction method to a 100 deg$^2$ region ($209.75\deg \leq l \leq 219.75\deg$ and $|b| \leq 5\deg$) in the third Galactic quadrant, which was mapped as part of the Milky Way Imaging Scroll Painting (MWISP) project, covering three CO isotopologue lines, and  derived distances and masses for 11 molecular clouds, including the Maddalena molecular cloud and Sh 2-287. The results indicate that the distance of the Perseus Arm in this region is about 2.4 kpc and molecular clouds are present in the interarm regions.  
\end{abstract}




\section{Introduction} \label{sec:intro}



Molecular cloud distances are essential to the study of star formation \citep{2007ARA&A..45..565M,2015ARA&A..53..583H,2015AJ....150..147F,2018ARA&A..56...41M} and the  Galactic structure \citep{2001ApJ...547..792D, 2016SciA....2E0878X,2018RAA....18..146X}. They are raw materials from which new stars form \citep{2012ARA&A..50..531K} and are spiral arm tracers \citep{2001ApJ...547..792D,2013ApJ...772..107D}. In both kinds of studies,  determining distances to molecular clouds is extremely important, particularly in the derivation of intrinsic physical properties, such as mass and size, and in the depiction of Galactic spiral arms.

However, molecular cloud distances are usually hard to obtain, due to the absence of reliable distance indicators, such as stellar luminosity. Principally, we have six approaches to obtain molecular cloud distances: (1) comparison of star counts between obscured  and non-obscured fields \citep{1923AN....219..109W,1937dss..book.....B, 1986A&A...168..271M,2012ApJ...751..157F}; (2) kinematic distances with Galactic rotation curves \citep{2014ApJ...783..130R}; (3) trigonometric or photometric distances of nearby OB-associations \citep{1978ApJS...38..309H,1992A&AS...94..211G, 1999AJ....117..354D, 2003A&A...404..519P}, young open clusters  \citep{2018A&A...618A..93C}, and \HII\ regions \citep{1976A&A....49...57G, 2007A&A...470..161R}; (4) trigonometric distances of masers in molecular clouds \citep{2006Sci...311...54X, 2009ApJ...693..397R, 2013ApJ...775...79Z, 2019AJ....157..200Z}; (5) trigonometric parallaxes of young stellar objects (YSOs) in molecular clouds \citep{2019MNRAS.487.2522M}; (6) identifying jump positions of stellar optical extinction caused by molecular clouds along the line of sight \citep{2014ApJ...786...29S,  2019ApJ...879..125Z, 2019A&A...624A...6Y}. The first method, star count, however, suffers from large uncertainties in stellar density distribution and variations of the stellar luminosity function. For the second method, the uncertainties ($\sim$0.7 kpc) of kinematic distances are large, and this method is not applicable to local molecular clouds and velocity crowding regions, e.g., in the Galactic anticenter direction. In addition, in the inner Galaxy, this approach has to deal with the distance ambiguity problem \citep{2003ApJ...582..756K}. The third method relies on indirect tracers and may have large systematic errors, and it is only useful for star-forming regions, particularly high-mass star-forming regions. In the fourth method, masers are not present in all molecular clouds and can only provide distances for single points, which may slightly deviate from molecular cloud main bodies \citep{2012ApJ...751..157F,2018ApJ...869...83Z}. In the fifth method, YSOs usually are deeply embedded inside molecular clouds and the high optical extinction makes them hard to observe, so the YSO parallax method is only suitable to derive distances for  nearby star-forming regions. Traditionally, the sixth method  was limited by large uncertainties in stellar distances and extinctions \citep{1992ApJS...83..273P}  and was unable to perform robust statistical analysis. However, this situation was changed by the release of \textit{Gaia} DR2 \citep{2016A&A...595A...1G, 2018A&A...616A...1G}.

The \textit{Gaia} DR2 catalog has enabled us to derive distances of many molecular clouds \citep{2019ApJ...879..125Z,2019A&A...624A...6Y}, as well as their structures along the line of sight \citep{2018A&A...619A.106G}. \textit{Gaia} DR2 contains about 60 million stars that have both $G$-band extinction (\ag) and parallax ($>$ 0.2 mas and relative errors $<$ 20\%) measurements, which is capable of revealing molecular cloud distances. In the Galactic plane ($|b|<5\deg$), the average surface density of \textit{Gaia} stars is about 3670 deg$^{-2}$. Due to the large uncertainties, \ag\ should be used statistically \citep{2018A&A...616A...8A}. Usually, only \textit{Gaia} DR2 parallaxes  are  used and the extinction is derived with the aid of photometric observations at multiple wavelengths \citep{2018ApJ...869...83Z,2019ApJ...879..125Z}, but \citet{2019A&A...624A...6Y} showed that \ag\ in the \textit{Gaia} DR2 catalog is also able to derive reliable molecular cloud distances. Most molecular clouds that have distances determined are local and at high Galactic latitudes, where dust environments are relatively clean. Models that work fine at high Galactic latitudes are likely to fail at low Galactic latitudes.  For instance, \citet{2019ApJ...879..125Z} found that the ``ramp'' in foreground extinction would make derived distances systematically underestimated.

In addition to molecular cloud distances, \textit{Gaia} DR2 data have been intensively used to derive the three dimensional (3D) dust map within about 3 kpc of the Sun \citep{2019MNRAS.483.4277C,  2019arXiv190105971L, 2019A&A...625A.135L,2019arXiv190502734G}, which may be used to estimate molecular distances. For example, in a most recent work, \citet{2019arXiv190502734G} derived a 3D dust distribution ($\delta >  - 30$\deg ) with \textit{Gaia} DR2 parallaxes and stellar photometry measurements of Pan-STARRS 1 \citep{2016arXiv161205243F} and 2MASS \citep{2006AJ....131.1163S}. However, the radial pattern in 3D dust maps indicates that uncertainties in stellar distances or extinctions are still large.  Furthermore,  dust has no radial velocity measurements and is present in both molecular and atomic gases. Consequently, deriving molecular cloud distances from 3D dust maps is not straightforward.



In this work, we make new attempts to calculate molecular cloud distances in the Galactic plane, where the complexity of dust distribution along the line of sight  makes the variation of \ag\ irregular and difficult to model.  In order to correctly calibrate the \ag\ of stars in the direction of molecular clouds (on-cloud stars), we argue that the irregularities in variations of on-cloud star \ag\ would also be seen in the \ag\ of  stars around molecular clouds (off-cloud stars), providing that the  distribution of diffuse dust is roughly uniform over the on- and off-cloud regions (as would be the case wherever the molecular clouds themselves are not exceedingly large). This pattern of irregularity in the extinction can be first adequately traced in the direction of nearby, off-cloud stars and then subtracted from that of on-cloud stars.

We describe this new baseline subtraction method and apply it to a region ($209.75\deg \leq l \leq 219.75\deg$ and $|b| \leq 5\deg$) observed by the Milky Way Imaging Scroll Painting (MWISP) survey, a CO mapping project toward the northern sky. This region contains three prominent components: 1) the giant molecular cloud (GMC) G216-2.5 \citep{1985ApJ...294..231M},  2) Sh 2-287 \citep{2016A&A...588A.104G}, and 3) Sh 2-284 \citep{2010A&A...509A.104D,2011MNRAS.410..227C}.  An alternative name of G216-2.5 (hereafter G216) is the Maddalena molecular cloud, whose distance is about 2.2 kpc \citep{1991ApJ...379..639L} from the Sun, containing 1-6$\times10^5$ \msun\ of molecular gas \citep{1994ApJ...432..167L}. Sh 2-287 (hereafter S287) is an \HII\ region that is ionized by an O9.5 star \citep{1984NInfo..56...59A}, and it appears to be associated with G216 \citep{1994ApJ...432..167L}. S287 displays a filamentary structure \citep{2013ApJ...772...45E,2014ApJ...788....3E,2016A&A...588A.104G} and contains many star formation activities \citep{2005A&A...442L..57V}.  Sh 2-284 (hereafter S284) is an \HII\ region ionized by the open cluster Dolidze 25 in the Milky Way's Outer Arm. Recently,  \citet{2015A&A...584A..77N} estimated the distance and age of Dolidze 25, which are $\sim$4.5 kpc and $<$ 3 Myr respectively, consistent with previous works \citep{1993AJ....105.1831T, 2009A&A...503..107P, 2010A&A...509A.104D, 2011MNRAS.410..227C}.

We begin the next section (\S\ref{sec:data}) with descriptions of the MWISP observations and \textit{Gaia} DR2 data reductions. In \S\ref{sec:result}, we describe the survey results, principles of distance calculation using the MWISP survey data and \textit{Gaia} DR2 stars, and the molecular cloud  distance catalog.  Discussions about the method and derived distances are presented in \S\ref{sec:discuss}, and we summarize the conclusions in \S\ref{sec:summary}. 




{\catcode`\&=11 \gdef\BolattoHeyer{\citep{2013ARA&A..51..207B, 2015ARA&A..53..583H}}}
{\catcode`\&=11 \gdef\Schoier{\citep{2005A&A...432..369S}}}

\begin{deluxetable*}{lccccccc}
\tablecaption{Observation parameters of the CO lines.\label{Tab:lineParameters}}
\tablehead{
\colhead{CO line} & \colhead{Rest frequency} & \colhead{Critical density\tablenotemark{a}} &\colhead{HPBW} & \colhead{$T$$_{\rm sys}$} & \colhead{$\eta_{\rm mb}$} &
\colhead{$\delta v$} & \colhead{noise rms} \\
\colhead{ $(J=1\rightarrow0)$ } & \colhead{(GHz)} & \colhead{($10^3$ cm$^{-3}$)} &\colhead{($''$)} & \colhead{ (K) } & \colhead{} &
\colhead{(\kms)} & \colhead{(k) } 
}

\startdata
\cofs  &   115.271204&  $\sim$0.06 & 49 &220-300  &40\%  & 0.16 & 0.5 \\\
\coss &  110.201353  & $\sim$6 & 51    &140 -190  &50\%  & 0.17&  0.25 \\
\cots &  109.782183  &  $\sim$18  &   50    &140 -190  &   50\% & 0.17& 0.25\\
\enddata
\tablenotetext{a}{The critical density is calibrated with optical depths (radiative trapping) based on the value of 2,000 cm$^{-3}$ \BolattoHeyer. The Einstein-$A$ coefficient of \cofs\ is $7.2\times10^{-8}$, and $6.3\times10^{-8}$ for the other two lines \Schoier.   The typical optical depths of \coss\ and \cots\ are $\sim$0.3 \citep{2015AJ....150...60L} and $\sim$0.1 \citep{2016AJ....152..117Y}, respectively, while the optical depth of \cofs\ is $\sim$30, estimated with the \cofs/\coss\ ratio \citep[$\sim$100, ][]{2003ApJ...598.1038G}.   }    
\end{deluxetable*}






\section{Observations and data reduction}
\label{sec:data}

\subsection{The MWISP survey}

 Observations of the MWISP\footnote{\href {http://www.radioast.nsdc.cn/mwisp.php}{http://www.radioast.nsdc.cn/mwisp.php}} survey contain three CO isotopologue lines, \cof, \cos, and \cot. The MWISP project is a new uniform and large-scale CO survey of the Galactic plane ($-10\deg \leq l \leq 250\deg$ and $|b| \leq 5\deg$), which has refined and newly detected many molecular clouds \citep{2015ApJ...798L..27S,2017ApJS..230...17S,2017ApJS..230....5W,2017ApJS..229...24D,2018ApJS..238...10L}. We refer the reader to \citet{2019ApJS..240....9S} for a more detailed description of the  MWISP survey and here briefly review the observation and data reduction processes.

Observations were conducted with the Purple Mountain Observatory (PMO) 13.7m millimeter telescope at Qinghai station, from June 2012 to April 2016. The PMO 13.7m telescope is equipped with a 3$\times$3 array Superconducting  Spectroscopic Array Receiver (SSAR) working in sideband separation mode \citep{2012ITTST...2..593S}. The 1-GHz-bandwidth fast Fourier transform spectrometer (FFTS), which performs spectral analysis, has 16384 channels, resulting a velocity resolution of $\sim$0.16 \kms\ at 115 GHz. The system temperature is approximately 220-300 K for the upper sideband (\cofs) and about 140-190 K for the lower sideband (\coss\ and \cots). The beam HPBW  is $\sim$51\arcsec\ at 110.2 GHz with a $\sim$5\arcsec\ pointing uncertainty, and the main beam efficiency is approximately 40\% for \cofs\ and 50\% for \coss\ and \cots, according to the status report\footnote{\href {http://www.radioast.nsdc.cn/ztbg/ztbg2015-2016engV2.pdf}{http://www.radioast.nsdc.cn/ztbg/ztbg2015-2016engV2.pdf}} of the PMO telescope.



The whole region was divided into $30'\times 30'$ individually mapped tiles, although in practice the survey area was chosen to be slightly larger than the tile sizes in order to reduce edge effects. The observations were carried out in a position-switch On-The-Fly (OTF) model with a scan rate of $\sim$50\arcsec\ $\rm s^{-1}$, and the scan row interval is 10\arcsec. We processed the spectral lines with the GILDAS \citep{2005sf2a.conf..721P} software and regridded the data into pixels of $30''\times 30''$. All the tiles were collectively mosaicked into FITS cubes after baseline calibrations. The single-channel  noise of the  \cofs\ line was $\sim$0.5 K (\cofs) at a velocity resolution of 0.16 \kms, and $\sim$0.25 K (for the \coss\ and \cots\ lines) at 0.17 \kms. Table \ref{Tab:lineParameters} summarizes the observation parameters of the three CO spectral lines.

\subsection{\textit{Gaia} DR2}

The reduction of \textit{Gaia} DR2 follows the procedure of \citet{2019A&A...624A...6Y} with slight modifications. We require that \ag\ $>0$ and the ratio of parallax errors (\paraerr) to  parallaxes (\para) are less than 20\% (i.e., \paraerr/\para\ $<$ 0.2 ). When \paraerr/\para\ $\geq$ 20\%, the distances (\dis) derived from \para\ is unreliable \citep{2015PASP..127..994B}. We took the reciprocal of  20,000 samples from the parallax normal distribution $\mathcal{N}(\varpi, \Delta \varpi )$  to derive the mean and standard derivation of $D$. The standard deviation of \ag\  is calculated with 
\begin{equation}
\Delta A_G =  \frac{1}{2}\left( A_G^{\mathrm{upper}} - A_G^{\mathrm{lower}} \right),
\end{equation}
where  $ A_G^{\mathrm{lower}}$ and $A_G^{\mathrm{upper}}$  are the 16th and 84th percentiles respectively, which are given by the \textit{Gaia} DR2 catalog.

For each molecular cloud region, \textit{Gaia} DR2 stars are further classified using footprints of molecular clouds in data cubes and CO integrated intensity thresholds, details of which are described in \S\ref{sec:method}.


\section{Results}
\label{sec:result}

In this section, we present our new procedure of calculating molecular cloud distances, as well as the results of applying it to the CO map of a chosen 100 deg$^2$ region containing 31 molecular clouds. Of the three CO isotopologue lines, \cofs\ and \coss\ were well mapped, but the signal of \cots\ is too faint so it was only detected in S287 \citep[see figure 10 of][]{2016A&A...588A.104G}. Consequently, we only display the maps of \cofs\ and \coss\ and ignore \cots, and for distance calculations, \cofs\ maps alone are sufficient.

 \begin{figure*}[ht!]
\gridline{\fig{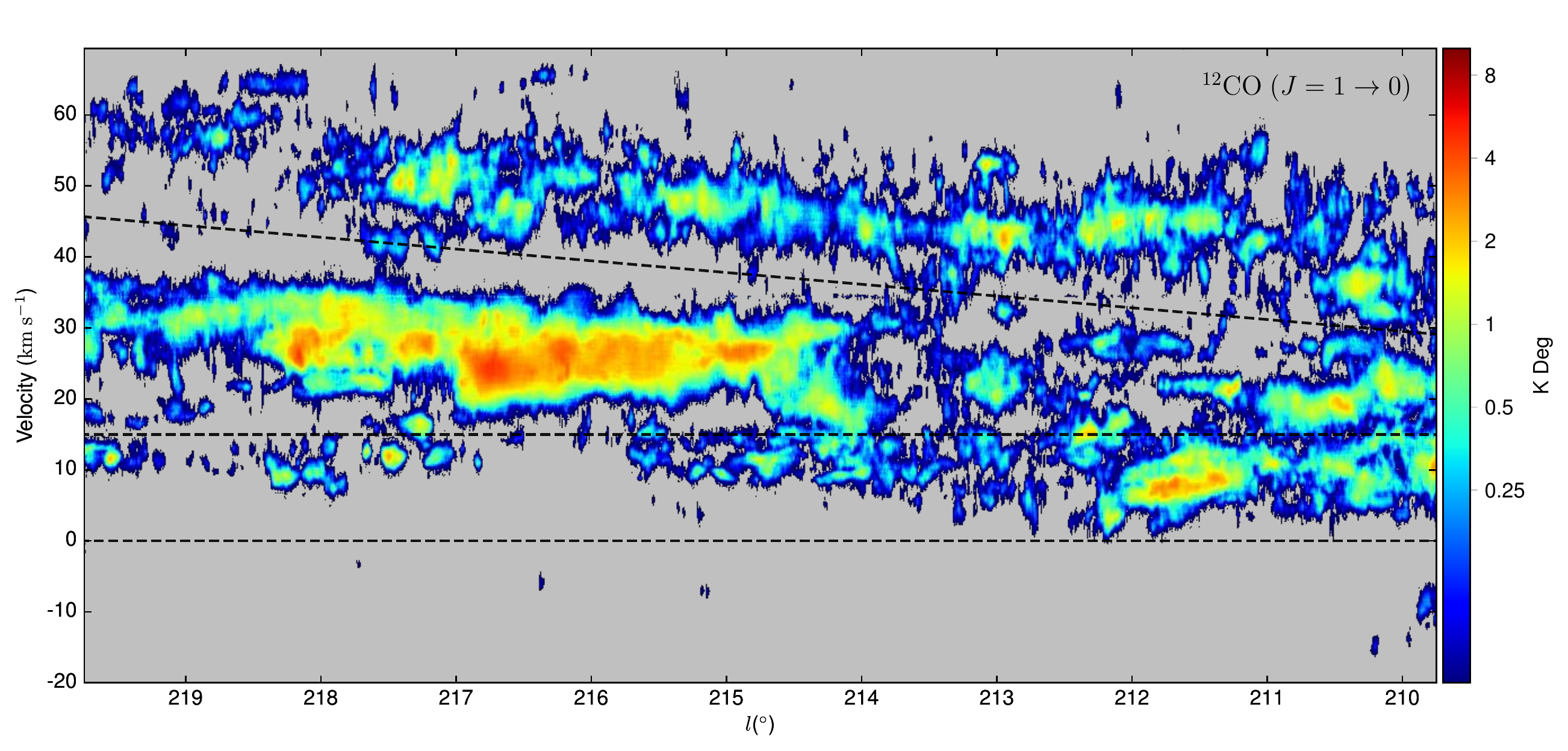}{0.9\textwidth}{(a)} 
          }
\gridline{\fig{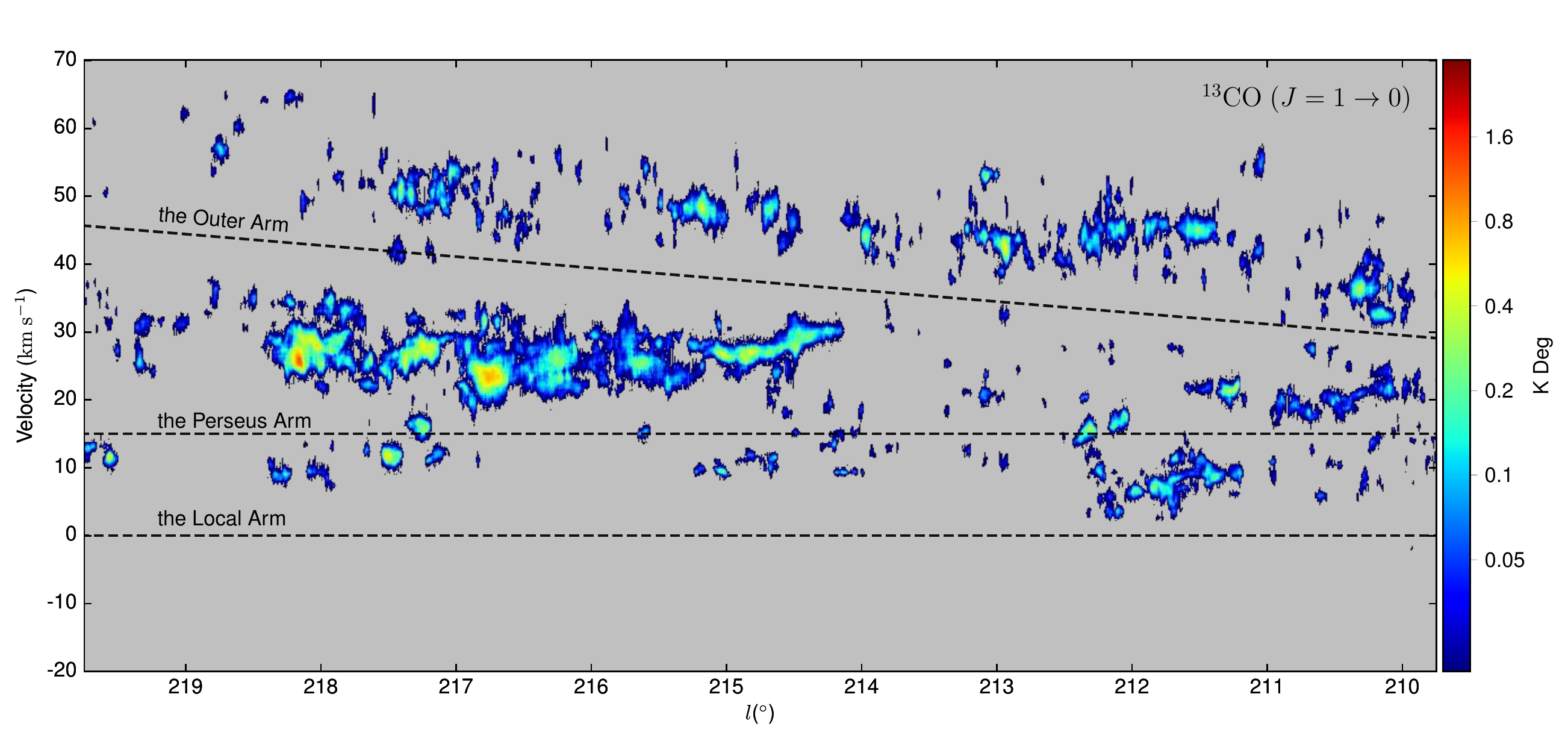}{0.9\textwidth}{(b)} 
          }
\caption{  The Galactic longitude-velocity ($l$-$V$) diagram of \cofs\ (a) and \coss\ (b). The dashed horizontal lines mark the dividing velocities between arms, $V$=15 \kms\ for the Local the Perseus Arm and $V$= $(1.656\times l-318.196)$ \kms\ for the Perseus and Outer Arm. In panel (a), the faint horizontal line at 34 \kms\ is due to a bad velocity channel. \label{fig:ppvg210} }
\end{figure*}
 
\subsection{The CO data}

In order to see the distribution of molecular clouds with respect to the radial velocity and the Galactic longitude, we integrated the spectra along the Galactic latitude and display the Galactic longitude-velocity ($l$-$V$) diagram in Figure \ref{fig:ppvg210}. To avoid the signal being overwhelmed by the noise, we used the  source finder software \texttt{Duchump} \citep{2012MNRAS.421.3242W} to mask the signal range in \coss\ and \cofs\ spectra before integrating. \texttt{Duchump} misses part of  weak CO emissions ($<3 \sigma$), but it was only used to produce $l$-$V$ diagrams and to integrate  \cofs\ and \coss\ intensity maps, not any part of subsequent molecular cloud identifications and distance calculations. We split \cofs\ and \coss\ data cubes into three parts: the Local Arm, the Perseus Arm, and the Outer Arm. The Local Arm is defined as the velocity range [0, 15] \kms.  Because the boundary of the Perseus and Outer Arm is not horizontal in the $l$-$V$ diagram, we drew a tilted line between these two arms with the supporter vector machine (SVM) algorithm supervised by the clouds that appear to be in the Perseus or the Outer Arms, producing an $l$-$V$ relationship of $V$= $(1.656\times l-318.196)$ \kms. This division is very rough and is only used as a first order approximation. 



\begin{figure*}[ht!]
\plotone{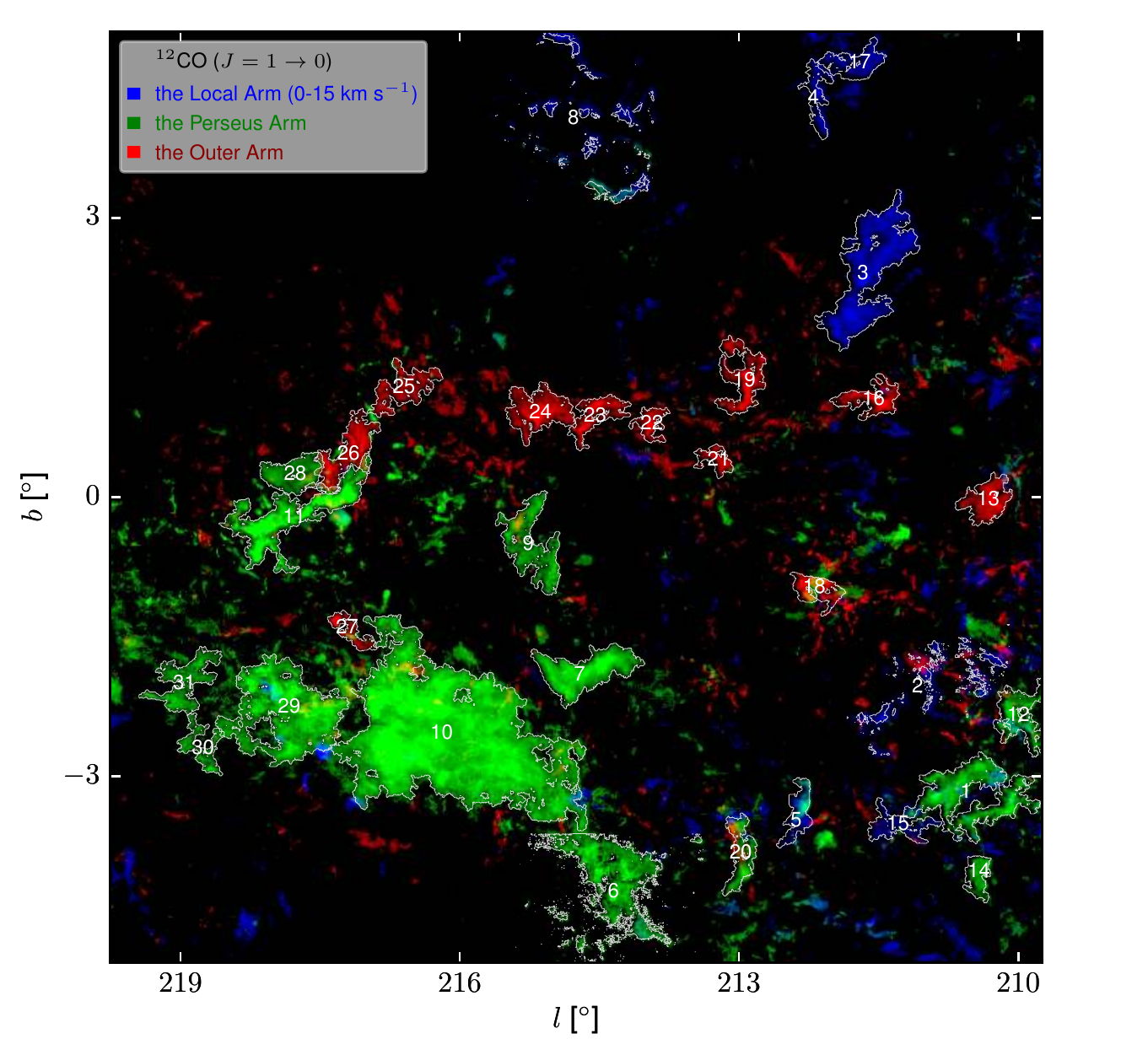}
\caption{The three-color image of \cof. The integration interval in the $l$-$V$ space for the Local Arm (blue), the Perseus Arm (green), and the Outer Arm (red) are delineated in Figure \ref{fig:ppvg210} with dashed black lines. The areas of 31 molecular clouds (see Table \ref{Tab:cloudDis}) are marked with white contours. For molecular clouds identified with dendrogram, masked areas are shown, while for those identified visually, the contour level shown is 4 K \kms.        \label{fig:co12rgb}}
\end{figure*}

\begin{figure*}[ht!]
\plotone{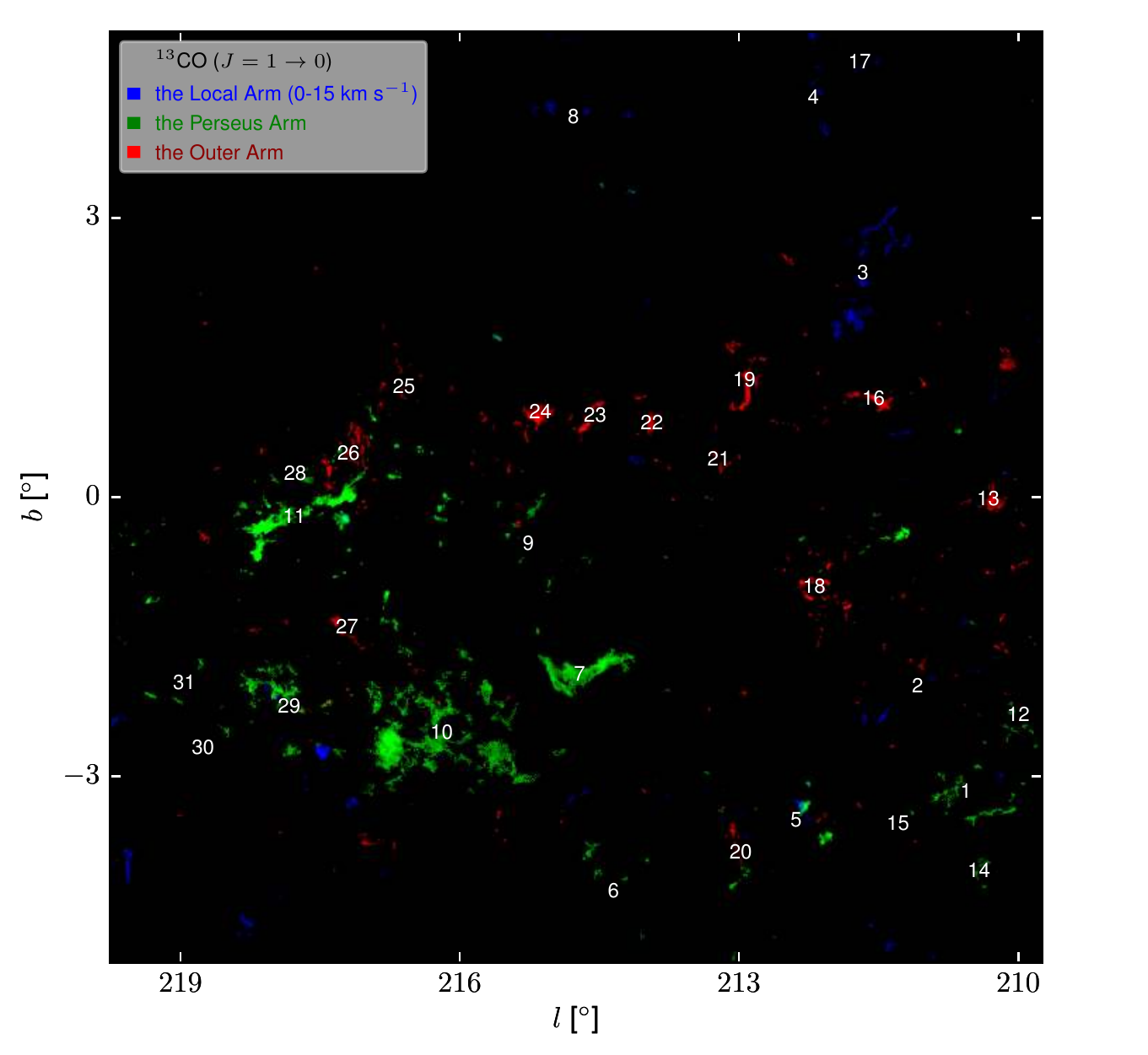}
\caption{Same as Figure \ref{fig:co12rgb} but for \cos. \label{fig:co13rgb}}
\end{figure*}

Figure \ref{fig:co12rgb} and \ref{fig:co13rgb} show the integrated  \cofs\ and \coss\ intensity maps, color coded by Galactic Arm segments. The most prominent component is G216 in the Perseus Arm (green), at about 25 \kms.  Due to their relatively near distances, molecular clouds in the Local Arm (blue) are more dispersed than that of the other two arms, and tend to be at high Galactic latitudes \citep[also see the CO surveys toward the Canis Minor region done by][]{2001ApJ...547..792D}. Interestingly, the primary component of the Outer Arm (red) is roughly parallel to the Galactic plane but with a slight offset ($\sim$0.8\deg), about 63 pc at a distance of 4.5 kpc \citep{2015A&A...584A..77N}.

\subsection{Distances}
\label{sec:arm}

\subsubsection{Methodologies}
\label{sec:method}

We refined the approach of \citet{2019A&A...624A...6Y}  to calculate distances of molecular clouds in the Galactic plane using the \textit{Gaia} DR2 catalog. Molecular clouds increase the \ag\ of background stars along the line of sight,  and  ideally,  the \ag\ of foreground and background stars approximately follows Gaussian distributions but with different means and standard deviations. Because the \ag\ in \textit{Gaia} DR2 is truncated,  \citet{2019A&A...624A...6Y} used a model of two switching truncated Gaussian distributions to detect the switching positions (i.e., distances of molecular clouds) in on-cloud star \ag\ for about 50 molecular clouds, mostly of which are at high Galactic latitudes.
 
For molecular clouds in the Galactic plane, the contamination of foreground dust makes the distribution of \ag\ along the line of sight complicated, in which case the  model of \citet{2019A&A...624A...6Y} is inapplicable. However, we argue that the \ag\ of off-cloud stars (around molecular clouds) can be used to calibrate the on-cloud \ag, which is supported by our finding that after calibration, the on-cloud  star \ag\ is approximately Gaussian, allowing us to identify the \ag\ due to specific molecular clouds with variations that remain after subtracting off this baseline. The resulting baseline-subtracted \ag\ are more appropriately modelled with full Gaussian distribution.

\begin{figure*}[ht!]
\gridline{\fig{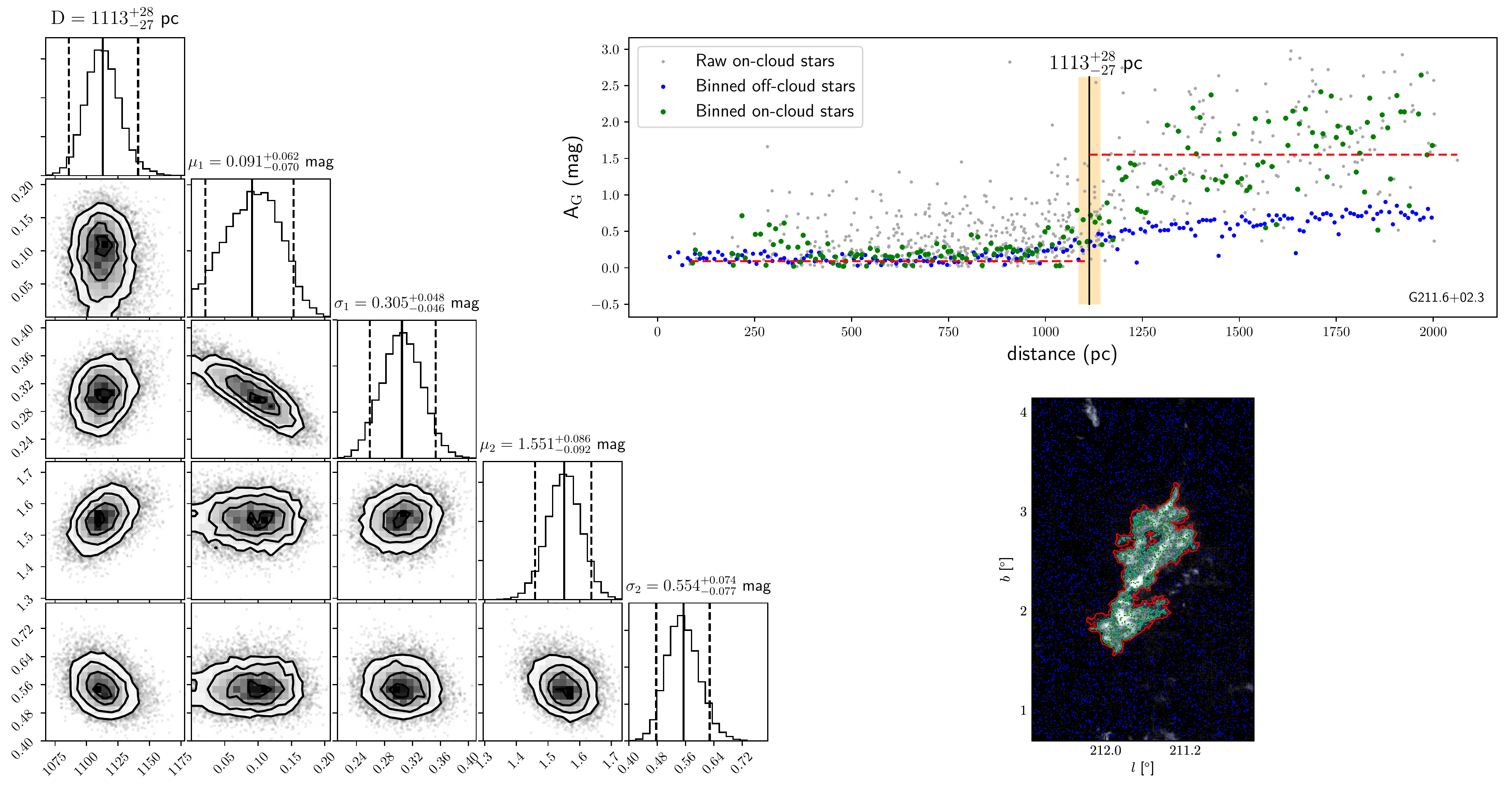}{0.88\textwidth}{(a)} 
          }
\gridline{\fig{{G211.6+02.3DistanceAG}.pdf}{0.88\textwidth}{(b)} 
          }
   
\caption{The distance of G211.6+02.3 using (a) the method of \citet{2019A&A...624A...6Y}  and  (b) with the baseline subtraction method. In the bottom right panel of (a) and top middle panel of (b), the green and blue dots represent on- and off-cloud stars, and in order to keep the plots from becoming too cluttered, we only draw up to 3000 randomly chosen stars for both on- and off-cloud populations. The red and blue contours refer to the footprint (produced by dendrogram) and the signal level used to classify \textit{Gaia} stars.   In (a) and (b), the shadowed area marks the 95\% highest posterior density (HPD) interval of the distance, and corner plots of MCMC samples of five modeled parameters (and their 95\% HPD) are displayed on the left panels. We calculated 10 chains, each of which contains 1000 thinned samples (every 15) with extra 50 burn-in (the first 50 thinned samples were removed). In the right plots, the green and blue points are the binned on- and off-cloud stars, respectively, but they are only used for visual confirmation and not for calculation. The bottom right panel of (b) displays the distribution of calibrated on-cloud star values of  \ag.  \label{fig:g211dis}} 
\end{figure*}

The basic assumption of this method is that the distribution of diffuse dust over the on- and off-cloud regions is approximately uniform. The validity of this assumption naturally changes from region to region, but we argue that the \ag\ caused by diffuse dust in the on- and off-cloud regions is approximately equal at least in the third Galactic quadrant, as shown by the foreground \ag\ (see Figure \ref{fig:g211dis}). 
 
To check the validity of our approach, we calculated the distance of a molecular cloud in the Local Arm, G211.6+02.3 ($l=211.627\deg$, $b=2.384\deg$, and $v= 7.4$ \kms), as demonstrated in Figure \ref{fig:g211dis}. Panel (a) of Figure \ref{fig:g211dis} displays the result derived with the truncated Gaussian model \citep{2019A&A...624A...6Y}, while panel (b) shows the results of the baseline subtraction method. The green and blue points are the averaged \ag\ and distances of on- and off-cloud stars (every 10 pc) weighted by their errors, respectively. The binned \ag\ is only used for visual  confirmation and is not any part of calculation, as neither approach requires binning. The \ag\ jump of G211.6+02.3 is clear and the foreground \ag\ is not complicated, so its distance  was able to be reliably calculated using with the method of \citet{2019A&A...624A...6Y}.  However, the variation of the  off-cloud \ag\ (blue points) shows that the \ag\ contains the contribution from dust in diffuse molecular or atomic clouds, and the baseline subtraction method removes unrelated \ag\ efficiently. Details of the baseline subtraction procedure are presented in the next section.

 \subsubsection{The Procedure}

\label{sec:procedure}

In this section, we describe the procedure of  calculating the distance to molecular clouds using the baseline subtraction method. Generally, the procedure includes four main steps: (1) identifying molecular clouds; (2) classifying on- and off-cloud stars; (3) fitting an \ag\ baseline with off-cloud stars; (4) calibrating on-cloud stars with the \ag\ baseline and finally calculating the distances. Compared with the method of \citet{2019A&A...624A...6Y}, the new approach adds an extra step of subtracting the \ag\ baseline from on-cloud stars and uses full (i.e., not truncated) Gaussian distribution instead.

First, we identify large-size \cofs\ clouds (\coss\ is not used for cloud identification) with the dendrogram algorithm \citep{2008ApJ...679.1338R}. The dendrogram algorithm maps data onto tree-like hierarchical structures, in which a leaf is a structure that has no child structures and a trunk is a structure that has no parent structures. The reason for choosing the  dendrogram algorithm is that it controls the merge size of molecular clouds. We use \verb|astrodendro|, an implementation of the algorithm in Python specifically designed for computing   dendrograms of astronomical data\footnote{\href {http://www.dendrograms.org/}{http://www.dendrograms.org/}}.  The  dendrogram algorithm has three important parameters:  \verb|min_value|, \verb|min_delta| and  \verb|min_npix|. \verb|min_value| is the minimum value to consider in data cubes, while \verb|min_delta| is the minimum height of a structure against neighboring structures.  Both \verb|min_value| and \verb|min_delta| are assigned to 3$\sigma$ (1.5 K for \cofs). The third parameter is the minimum number of voxels contained in a leaf. Considering a 0.25-deg$^2$ molecular cloud that has a velocity dispersion of 1 \kms, its number of voxels is about 18,000, and therefore, we used a value of 10,000 for \verb|min_npix| to make the catalog complete. For distance calculations, usually only trunks are used, but the Maddalena cloud is too large, and we split it into two smaller regions. In total, \verb|astrodendro| found  28 molecular clouds. However, the algorithm misses molecular clouds that are too loosely distributed. Therefore, we searched  the cube by eye and identified three molecular clouds. Consequently, the final catalog contains 31 molecular clouds, and the sky projections of those molecular clouds are delineated in Figure \ref{fig:co12rgb}. 

This definition of molecular clouds hinges on line profile sensitivities, and lower noises would yield larger molecular cloud areas. Those molecular clouds with low \cofs\ emission ($<3\sigma$) are left out as off-cloud regions. Numerical simulations \citep{2014MNRAS.441.1628S} suggest that the mass fraction of CO-dark molecular gas is $\sim$40\%, and observations \citep{2014A&A...561A.122L} show that this fraction could be higher in diffuse molecular clouds (no \cofs\ emission) and lower in  \coss\ clouds. However, both the $\rm H_{2}$ column density ($\sim9\times10^{20}~ \rm cm^{-2}$) and visual extinction ($\sim$1 mag) of CO-dark molecular clouds is lower than that \citep[1-3 mag, ][]{2015ARA&A..53..583H} of CO molecular clouds, and as discussed at the end of this section, the effect of CO-dark molecular clouds is insignificant in distance calculations.




Secondly, stars are classified as either on- or off-cloud using the footprint of molecular clouds and CO integrated intensity thresholds. The footprint produced with \verb|astrodendro| is 3D, and we projected the 3D footprint onto the sky to obtain 2D masks of on-cloud regions (see the red contours in Figure \ref{fig:g211dis}). The raw on-cloud stars are the \textit{Gaia} stars that within the masked area. To obtain raw off-cloud stars, we double the lengths of the sides of the minimum rectangular region boxes (along the $l$ and $b$ directions) that contain the molecular clouds. \textit{Gaia} stars within the extended box region, but which lie outside of the masked areas are labeled raw-cloud stars. A maximum distance cutoff is set for both on- and off-cloud stars  to remove stars that are too far away. Near the edge of molecular clouds, many stars have low CO integrated intensities, while in the unmasked region, some areas show CO emission that is unrelated to the target CO molecular clouds. Consequently, we imposed CO integrated intensity  thresholds on the on- and off-cloud stars to further remove \textit{Gaia} stars that are unsuitable for distance calculations. On-cloud stars with CO integrated intensities  $<$ 5 K \kms\ (the signal level) or off-cloud stars with CO integrated intensities $>$1 K \kms\ (the noise level) were discarded. As discussed below, within acceptable ranges, the derived distances are insensitive to the choice of signal and noise levels, particularly the latter one.


Thirdly, we fitted the \ag\ baseline using off-cloud stars. Because \ag\   is  a monotonically increasing function of the distance, we fitted the baseline  using  monotonic (isotonic) regression. Before fitting, we sorted  \ag\ according to their distances, and then performed the isotonic regression weighted by the inverse-variance of \ag. The distance errors were ignored, because of the difficulty of performing isotonic regression using errors in both variables. We used a linear algorithm, the Pool Adjacent Violators Algorithm \citep[PAVA,][]{mair2009isotone}, which is implemented  in the Python  machine learning  package \texttt{sklearn} \citep{scikit-learn}, to perform the monotonic regression. We found that a small number of stars that have very high weights (standard deviations $<$ 0.05 mag) deform the fitted baseline seriously. Consequently, we set a lower cutoff (0.05 mag) for the standard deviation in the off-cloud stars. As discussed below, distances are insensitive to the exact threshold value chosen.   At the farthest end of off-cloud stars, the isotonic regression becomes unstable, and we extended off-cloud populations by including those in the next 300 pc range beyond the maximum distance cutoff to remove this edge effect.

Finally, we subtracted the \ag\ baseline from on-cloud stars, and derived the distance with Bayesian analysis and Markov chain Monte Carlo  (MCMC) sampling. After subtraction, the distribution of foreground and background \ag\ of on-cloud stars is regular and well described by Gaussian distributions as shown in the bottom right panel of (b) in Figure \ref{fig:g211dis}, indicating that the contamination has been largely removed. The model contains five parameters: the cloud distance ($D$), the extinction \ag\ ($\mu_1$) and standard deviation  ($\sigma_1$) of foreground stars, and the extinction \ag\ ($\mu_2$) and standard deviation ($\sigma_2$) of background stars. The likelihood of this altered model is similar to that of \citet{2019A&A...624A...6Y} except that the truncated Gaussian distributions are replaced with full Gaussian distributions. 

Here, we write out the likelihood formula for the new baseline subtraction method. The likelihood is calculated on the condition of a given star belonging to either foreground or background. For an on-cloud star, let the distance and \ag\ be $d_i\pm \Delta d_i$ and $A_{Gi}\pm \Delta A_{Gi}$, respectively, where $\Delta d_i$ and $\Delta A_{Gi}$ are standard deviations and $A_{Gi}$ is the calibrated on-cloud \ag. The probability of the star being in the foreground is 
\begin{equation}
f_i=\phi\left( \frac{ D- {d}_i }{ \Delta  {d}_i }  \right),
\end{equation}
where 
\begin{equation}
\phi\left(x\right)=  \frac{1}{\sqrt{2\pi}} \int_\infty^x e^{-t^2/2} \mathrm{d}t 
\end{equation}
 is the cumulative distribution function (CDF) of the standard normal distribution. The probability of being a background star is $(1-f_i)$.

The probability density function (PDF) of the Gaussian distribution is  
\begin{equation}
p\left(A_{G} |  \mu, \sigma\right) =   \frac{1}{ \sigma\sqrt{2\pi}} \exp\left( -\frac{1}{2}\left( \frac{A_{G}-  \mu}{  \sigma}    \right)^2 \right), 
\end{equation} 
where $\mu$ and $\sigma$ are the mean and standard deviation, respectively. For a foreground star, the likelihood of measuring $A_{Gi}$ is
\begin{equation}
PF_{i}=p\left(A_{Gi} |  \mu_1, \sqrt{\sigma_1^2+\Delta A_{Gi}^2} \right),
\label{equ:pf}
\end{equation}
while for a background star, the likelihood of measuring $A_{Gi}$ is
\begin{equation}
PB_{i}=p\left(A_{Gi} |  \mu_2, \sqrt{\sigma_2^2+\Delta A_{Gi}^2} \right).
\label{equ:pf}
\end{equation}
With above expressions, the likelihood of a star is 
\begin{equation}
p\left(A_{\mathrm{G}i} |  \mu_1, \sigma_1, \mu_2, \sigma_2,  D\right)=   f_i  PF_{i}+\left(1-f_i\right)PB_{i}.\end{equation} 
The total likelihood is the product of all on-cloud stars.

We solve the model with MCMC sampling. Due to the relative small number of background stars, the MCMC process occasionally converged to small molecular cloud  distances, taking almost all stars as background stars. In order to avoid this, we adjusted the prior uniform distribution of $D$ to have a minimum value located  200 pc farther than the nearest star in the population. That is, if the distance range of on-cloud stars is [$  D_{\mathrm{min}}$, $  D_{\mathrm{max}}$], the range of the prior uniform distribution of $D$  is [$ D_{\mathrm{min}}$+200 pc, $  D_{\mathrm{max}}$]. The priors of the other  four parameters were set to be  
\begin{equation}
\left\{
             \begin{array}{rcl}
  \rm \sigma_1  & \sim &\mathcal{E}\left(0.5\right), \\
\mu_1  &\sim& \mathcal{E}\left(0.5\right),\\
  \rm \sigma_2  & \sim &\mathcal{E}\left(\sigma_{50}\right), \\
\mu_2  &\sim& \mathcal{E}\left(\mu_{50}\right),\\
             \end{array}
\right.
\end{equation} 
where $\mathcal{E}$ represent an exponential distribution with mean as its only parameter (in units of magnitudes), and $\mu_{50}$ and $\sigma_{50}$ are the mean and standard deviation of the farthest 50 stars of on-cloud stars, respectively.

The MCMC algorithm used is the Gibbs sampler \citep{geman1984stochastic}. The form of transition probabilities are all Gaussian, and the standard deviation for $D$ is 100 pc and 0.5 mag for the other four parameters. For each parameter, we calculated 10 chains, with each chain containing 1000 thinned samples (every 15) with extra 50 burn-in (i.e., the first 50 thinned samples were discarded).

In Figure \ref{fig:g211dis}, as an example, we display the distance results for G211.6+02.3 and compare it with that  derived from the method of \citet{2019A&A...624A...6Y}. The method of \citet{2019A&A...624A...6Y} works on G211.6+02.3  because the foreground \ag\ is relatively simple, which is not the case for most molecular clouds in the survey region, where the foreground \ag\ cannot be modeled. Both methods give the same distance for G211.6+02.3 within errors, but the distance derived from the \ag\ of on-cloud stars is more  reliable, since the post-calibration distributions are closer to Gaussian. In addition, as shown by the corner map in panel (a), $\mu_1$ and $\sigma_1$ are strongly correlated in the approach of \citet{2019A&A...624A...6Y}, but are independent in the new method.  In Figure \ref{fig:g211dis}, the uncalibrated on-cloud \ag\ shows a slope near the molecular cloud position. This slope and the truncation of foreground \ag\ cause $\mu_1$ to be coupled with $\sigma_1$, and the removal of \ag\ baseline flats the on-cloud \ag, making the jump point sharp and decoupling $\mu_1$ from $\sigma_1$.

 As shown in Figure \ref{fig:g211dis}, both raw on- and off-cloud \ag\  grows slowly toward far distances. This tendency is still present in off-cloud \ag\ even if those stars near (within $\sim$16\arcmin) the on-cloud region were removed. We suspect that this rough  growth is due to diffuse dust in interarm regions (from 1 to 2 kpc), which can also be seen in the foreground \ag\ of molecular clouds in the Perseus Arm, such as Figure \ref{fig:g2152dis}, \ref{fig:g2162dis}, and \ref{fig:g2177dis}.  The removal of off-cloud stars near the edge of molecular clouds made the baseline  slightly flatter, but $\mu_2$ increased only by 0.01 mag. Consequently, for molecular clouds whose \ag\ jump points are clear, the effect of CO-dark molecular clouds on distances  are insignificant ($\sim$2 pc). However, in some extreme cases, it is possible that the visual extinction of CO-dark molecular clouds is high enough to diminish \ag\ jump points, causing failures in distance analyses.

\begin{figure*}[ht!]
\plotone{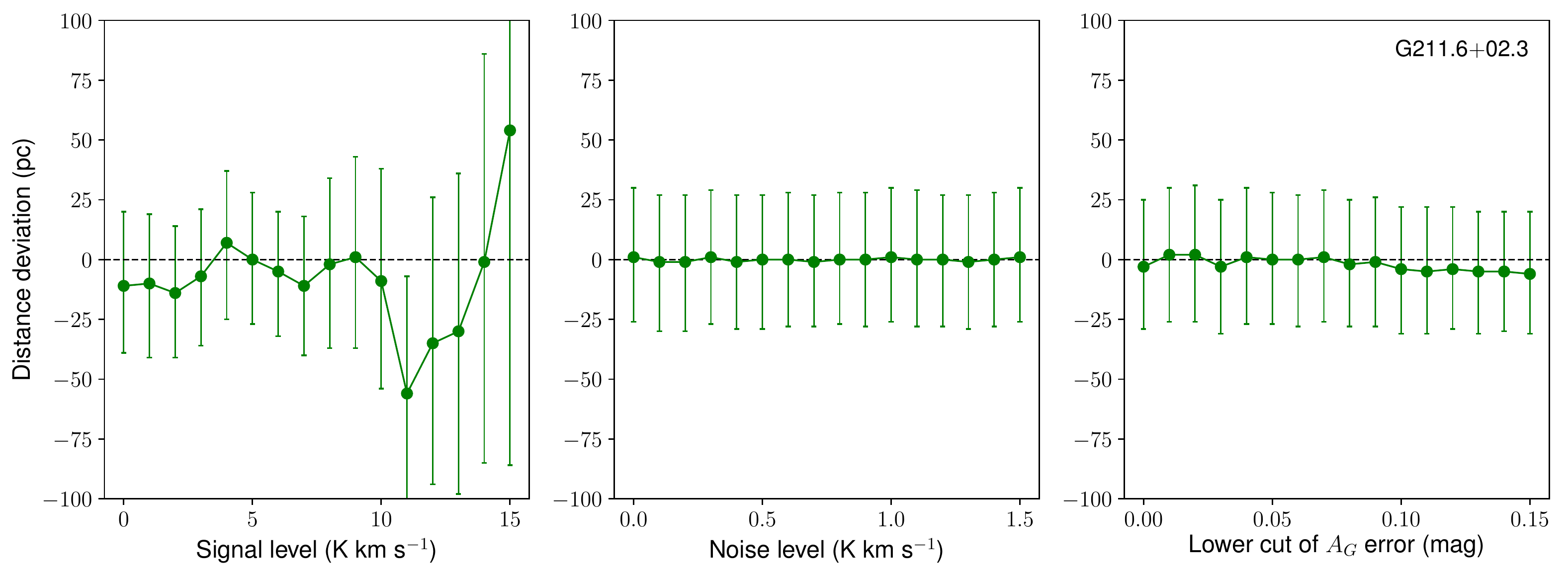}
\plotone{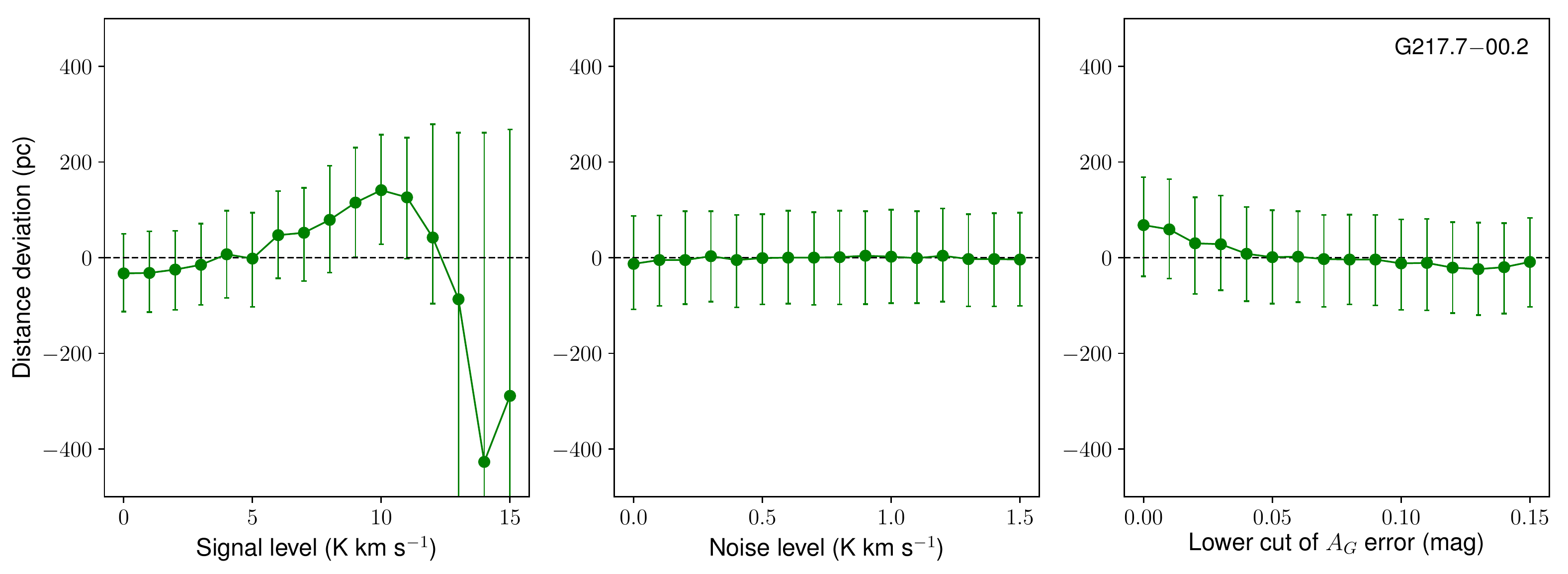}
\caption{Distance deviations of G211.6+023 and  G217.7-002 (S287) with parameter configurations. Three parameters, the signal level, the noise level, and the lower \ag\ error cutoff, were changed one at a time with respect to their reference values, which are 5 K \kms, 1 K \kms, and 0.05 mag, respectively. The reference distances of G211.6+023 and G217.7-002 calculated with these reference values are 1117 and 2190 pc, respectively. \label{fig:testpara}} 
\end{figure*}


\subsubsection{Distance results}

In this section, we present molecular cloud distances derived with this new baseline subtraction method. Before calculating distances, we examined the effect of parameter choices. \citet{2019A&A...624A...6Y} concluded that their derived distances contain 5\% systematic errors, which  are still present in the baseline subtraction method. Because we used off-cloud stars to produce baselines, the noise level (above which the off-cloud stars are removed)  and the minimum \ag\ error threshold (below which off-cloud stars are discarded) may affect the distances.

We examined three parameters: the noise level, the signal level (below which on-cloud stars are removed), and the \ag\ error threshold. Exploring the whole 3D parameter space is computationally expensive, so we changed one parameter at a time and kept the other two equal to their reference values. The reference values for the signal level, noise level, and the off-cloud \ag\ lower error threshold are 5 K \kms, 1 K \kms, and 0.05 mag, respectively. Figure \ref{fig:testpara} shows distance deviations of two molecular clouds, G211.6+023 and  G217.7-002 (S287). As can be seen in the figure, the distance is insensitive to the noise level and the lower \ag\  error cutoff, while the signal level affects the distance more prominently. When the signal level parameter is increased, the number of on-cloud stars decrease and the distance error becomes larger.  However, the distance is stable around 5 K \kms\ and the deviation is less than the 5\% systematic error. Consequently, we simply adopted the reference values for those three parameters, and for G211.1-02.1, a faint molecular cloud, a slightly lower signal level (4 K \kms) was used to include more on-cloud stars.

\begin{figure*}[ht!]
\plotone{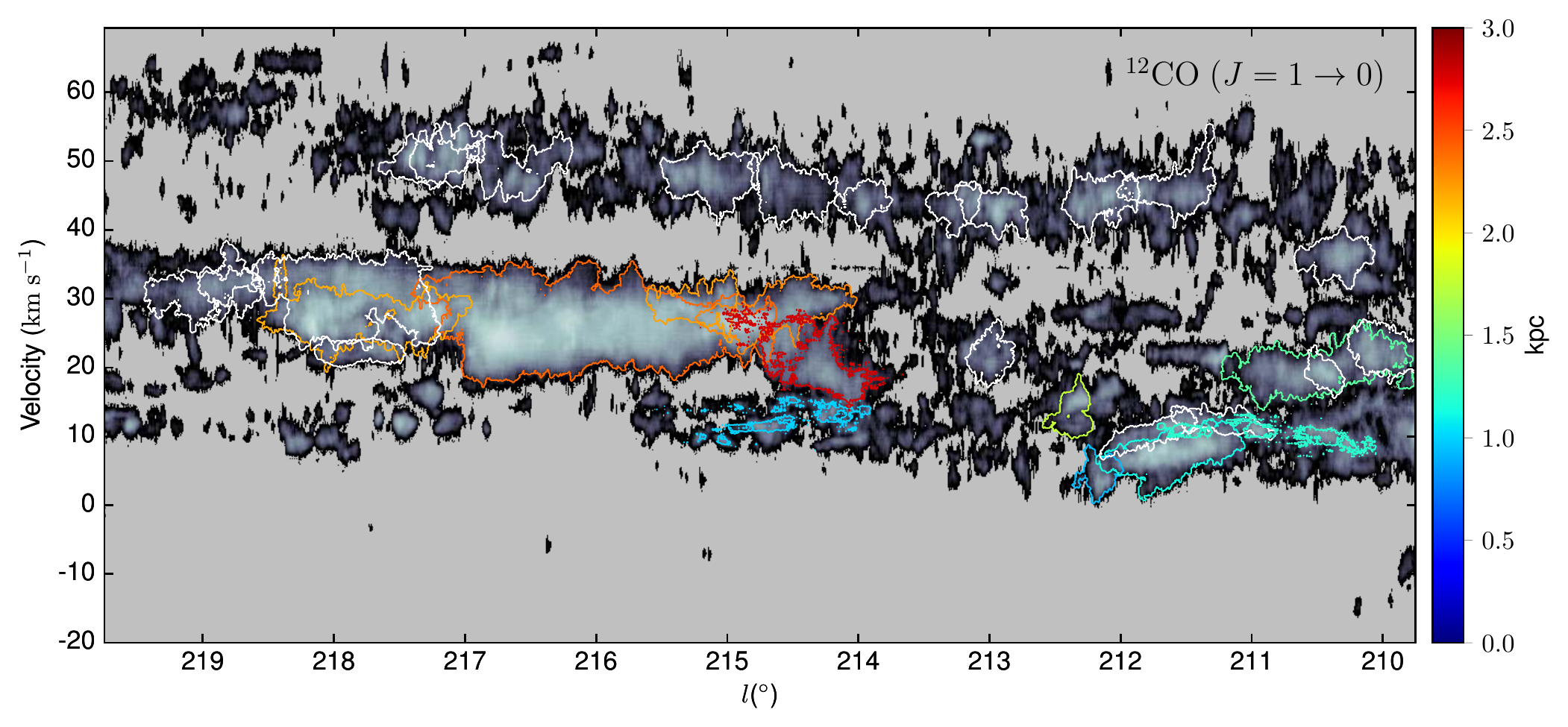}
\caption{Molecular cloud footprints and their distances.     The background is the $l$-$V$ diagram of \cofs\ (see Figure \ref{fig:ppvg210}), and distances are coded with colors (white for no distances). For molecular clouds identified with dendrogram, masked areas are shown, while for those identified visually, the contour level shown is 1 K deg.  \label{fig:footprint}}
\end{figure*}

With this baseline subtraction method, we performed distance calculations for 31 molecular clouds,  and the results are summarized in Table \ref{Tab:cloudDis}. The expected 5\% systematic error is not included. Eleven molecular clouds have well-determined distances, and we include the kinematic distances derived from the A5 model of \citet{2014ApJ...783..130R} for a comparison. In addition, we estimated the cloud mass by assuming a $^{12}$CO-to-H$_2$ mass conversion factor of X=$2.0\times 10^{20}$ cm$^{-2}$ K  km s$^{-1}$ \citep{2013ARA&A..51..207B}. Three possibilities may lead to erroneous distance calculations: (1) molecular clouds are too far ($>$ 3 kpc); (2) on-cloud stars are insufficient; (3) the values of \ag\ due to molecular clouds are too small. These reasons account for the 20 molecular clouds in Table \ref{Tab:cloudDis} for which distances were not calculated.  For 17 of those molecular clouds, we provide rough lower limits, nearer than which no jumpy positions are seen in the calibrated on-cloud \ag. The footprints of those 31 molecular clouds in the $l$-$V$ diagram are delineated in Figure \ref{fig:footprint}. The distance to G211.6+02.3 is displayed in panel (b) of Figure \ref{fig:g211dis}, while  the results for the other 10 molecular clouds are described in  Figures  \ref{fig:g2105dis} - \ref{fig:g2177dis}.

\begin{deluxetable*}{cccccccccccc}
\tablecaption{Distances of molecular clouds.\label{Tab:cloudDis}}
\tablehead{
\colhead{ID} &  \colhead{Name} & \colhead{$l$ } & \colhead{ $b$ } &\colhead{ $v_{\rm LSR}$} & \colhead{ Area }   & \colhead{ $D_{\rm Gaia}$\tablenotemark{a} } & \colhead{ N } & \colhead{Cutoff} & \colhead{Mass\tablenotemark{b}}     & $D_{\rm kinematic}$\tablenotemark{c} & Note \\
\colhead{   } & \colhead{   } & \colhead{(\deg)} & \colhead{(\deg)} &\colhead{(\kms)} & \colhead{deg$^2$} & \colhead{ (pc) }  &   & \colhead{ (pc) }& \colhead{ ($10^3$ \msun)}     & \colhead{ (kpc)}   
}
\startdata
  1  &  G210.5$-$03.1 &   210.524 &  -3.188 &    19.7  &  0.51 & $1373_{-  87}^{+  83}$ &   435 & 2000  &   9.3 & $1.97_{-0.74}^{+0.84}$ &         \\  
  2  &  G211.1$-$02.1 &   211.141 &  -2.103 &    10.2  &  0.50 & $1197_{-  93}^{+  88}$ &   220 & 2000  &   3.5 & $0.98_{-0.62}^{+0.69}$ & By eye\tablenotemark{d} \\  
  3  &  G211.6$+$02.3 &   211.627 &   2.384 &     7.4  &  0.67 & $1117_{-  30}^{+  26}$ &   677 & 2000  &   8.0 & $0.72_{-0.59}^{+0.66}$ &         \\  
  4  &  G212.1$+$04.2 &   212.161 &   4.275 &     4.0  &  0.10 & $ 931_{-  52}^{+  44}$ &    57 & 1500  &   0.7 & $0.42_{-0.42}^{+0.62}$ &         \\  
  5  &  G212.3$-$03.3 &   212.347 &  -3.345 &    14.2  &  0.11 & $1747_{- 167}^{+ 179}$ &   139 & 3000  &   3.4 & $1.32_{-0.64}^{+0.72}$ &         \\  
  6  &  G214.4$-$04.3 &   214.409 &  -4.305 &    21.6  &  0.75 & $2778_{- 212}^{+ 210}$ &  1361 & 4500  &  62.3 & $1.97_{-0.68}^{+0.76}$ & By eye\tablenotemark{d} \\  
  7  &  G214.6$-$01.8 &   214.673 &  -1.874 &    27.8  &  0.36 & $2301_{- 149}^{+ 142}$ &   738 & 3500  &  26.2 & $2.62_{-0.74}^{+0.84}$ & S287N   \\  
  8  &  G214.8$+$04.0 &   214.839 &   4.009 &    12.3  &  0.39 & $ 989_{-  58}^{+  61}$ &   186 & 2000  &   2.1 & $1.09_{-0.59}^{+0.65}$ & By eye\tablenotemark{d} \\  
  9  &  G215.2$-$00.5 &   215.224 &  -0.518 &    29.2  &  0.32 & $2221_{- 158}^{+ 168}$ &   590 & 3500  &  12.0 & $2.75_{-0.75}^{+0.84}$ &         \\  
 10  &  G216.2$-$02.5 &   216.216 &  -2.556 &    25.3  &  2.92 & $2411_{-  72}^{+  72}$ &  7407 & 3500  & 311.9 & $2.27_{-0.69}^{+0.76}$ & Maddalena \\  
 11  &  G217.7$-$00.2 &   217.800 &  -0.235 &    27.2  &  0.50 & $2190_{-  96}^{+ 100}$ &  1273 & 3500  &  46.0 & $2.40_{-0.68}^{+0.75}$ & S287    \\  
 12  &  G210.0$-$02.3 &   210.020 &  -2.357 &    22.6  &  0.20 &        $>$2500           &   279 & 3500  &   9.8 & $2.34_{-0.79}^{+0.90}$ &         \\  
 13  &  G210.3$-$00.0 &   210.343 &  -0.045 &    36.1  &  0.17 &        $>$3000           &   335 & 3500  &  32.1 & $4.17_{-1.02}^{+1.20}$ &         \\  
 14  &  G210.4$-$04.0 &   210.443 &  -4.039 &    18.9  &  0.09 &        $>$2000           &   172 & 3500  &   2.2 & $1.89_{-0.73}^{+0.83}$ &         \\  
 15  &  G211.3$-$03.5 &   211.313 &  -3.534 &    12.1  &  0.12 &          --            &    19 & 3500  &   0.8 & $1.15_{-0.64}^{+0.71}$ &         \\  
 16  &  G211.5$+$01.0 &   211.577 &   1.037 &    45.6  &  0.14 &        $>$2500           &   261 & 3500  &  48.6 & $5.57_{-1.19}^{+1.42}$ &         \\  
 17  &  G211.7$+$04.6 &   211.726 &   4.651 &    10.7  &  0.14 &          --            &   187 & 2000  &   1.0 & $1.02_{-0.62}^{+0.69}$ &         \\  
 18  &  G212.2$-$00.9 &   212.212 &  -0.987 &    44.2  &  0.11 &        $>$2500           &   226 & 3500  &  43.1 & $5.18_{-1.12}^{+1.32}$ & S284    \\  
 19  &  G212.9$+$01.2 &   212.966 &   1.239 &    42.8  &  0.21 &        $>$3000           &   380 & 3500  &  45.8 & $4.84_{-1.05}^{+1.23}$ &         \\  
 20  &  G213.0$-$03.8 &   213.002 &  -3.841 &    21.3  &  0.14 &          --            &   100 & 3000  &   3.7 & $2.01_{-0.70}^{+0.79}$ &         \\  
 21  &  G213.2$+$00.3 &   213.240 &   0.384 &    43.7  &  0.06 &        $>$3000           &   155 & 3500  &  13.6 & $4.92_{-1.05}^{+1.23}$ &         \\  
 22  &  G213.9$+$00.7 &   213.958 &   0.777 &    43.9  &  0.07 &        $>$2500           &   159 & 3500  &  16.7 & $4.85_{-1.02}^{+1.19}$ &         \\  
 23  &  G214.5$+$00.8 &   214.567 &   0.860 &    46.2  &  0.14 &        $>$3000           &   364 & 3500  &  40.7 & $5.11_{-1.04}^{+1.22}$ &         \\  
 24  &  G215.1$+$00.8 &   215.155 &   0.894 &    47.5  &  0.25 &        $>$3000           &   795 & 3500  &  78.0 & $5.23_{-1.05}^{+1.22}$ &         \\  
 25  &  G216.6$+$01.1 &   216.620 &   1.160 &    48.6  &  0.18 &        $>$3000           &   433 & 3500  &  40.1 & $5.17_{-1.00}^{+1.16}$ &         \\  
 26  &  G217.2$+$00.4 &   217.216 &   0.451 &    50.5  &  0.25 &        $>$3000           &   607 & 3500  &  76.2 & $5.37_{-1.02}^{+1.17}$ &         \\  
 27  &  G217.2$-$01.4 &   217.232 &  -1.421 &    50.1  &  0.08 &        $>$2000           &   110 & 3500  &  21.5 & $5.31_{-1.01}^{+1.16}$ &         \\  
 28  &  G217.7$+$00.2 &   217.789 &   0.235 &    22.3  &  0.19 &        $>$2000           &   359 & 3500  &   4.1 & $1.91_{-0.64}^{+0.70}$ &         \\  
 29  &  G217.8$-$02.2 &   217.853 &  -2.267 &    30.0  &  0.85 &        $>$3000           &  2654 & 3500  &  86.9 & $2.68_{-0.71}^{+0.78}$ &         \\  
 30  &  G218.7$-$02.7 &   218.779 &  -2.713 &    31.6  &  0.10 &        $>$2500           &   200 & 3500  &   6.1 & $2.80_{-0.71}^{+0.78}$ &         \\  
 31  &  G218.9$-$02.0 &   218.981 &  -2.018 &    31.5  &  0.22 &        $>$2500           &   604 & 3500  &  15.4 & $2.78_{-0.70}^{+0.78}$ &         \\  
\enddata
\tablenotetext{a}{The 5\% systematic error is not included. Lower limits are provided for 17 molecular clouds.}
\tablenotetext{b}{ The mass only takes account of CO-bright molecular gas. Gaia distances are used for cloud 1-11, while kinematic distances are used for cloud 12-31.   }
\tablenotetext{c}{Derived from the A5 model of \citet{2014ApJ...783..130R}.}
\tablenotetext{d}{These three molecular clouds were identified by eye.}
\end{deluxetable*}

\begin{figure*}[ht!]
\plotone{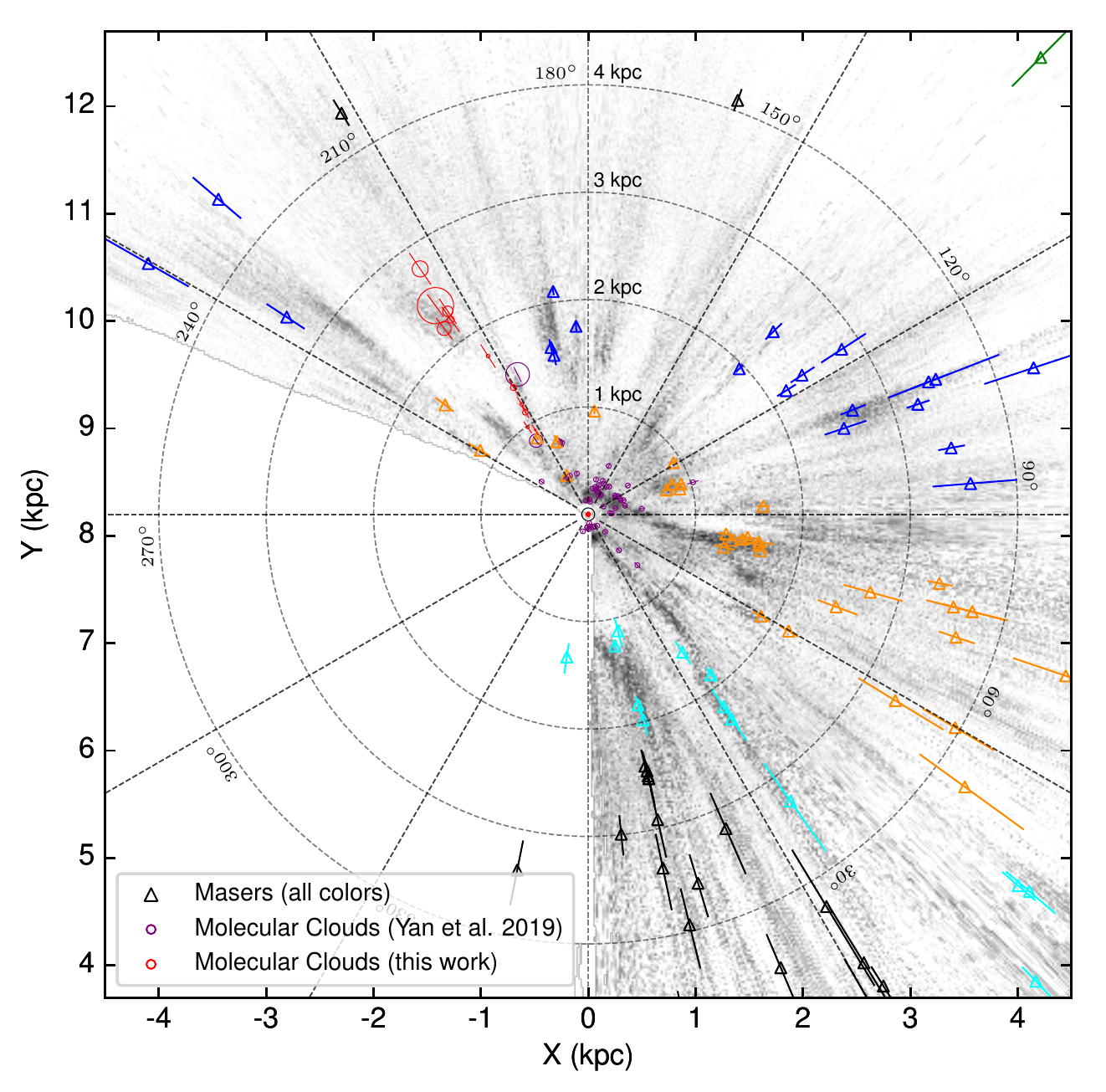}
\caption{Face-on view of  molecular clouds and masers. The background is an extinction map \citep{2019arXiv190502734G} of the Galactic plane (within a scale height of 200 pc), and the distance to the Galactic center is 8.2 kpc \citep{2019A&A...625L..10G}. 11 molecular clouds marked with red circles are from Table \ref{Tab:cloudDis}, and the sizes are  proportional to their masses. Masers are colored orange (the Local Arm), blue (the Perseus Arm), cyan (the Sagittarius Arm), green (the Outer Arm), and black (others). \citet{2019A&A...624A...6Y} determined distances for 52 molecular clouds (purple), most of which are local and at high Galactic latitudes.  In order to make a comparison, we rescaled the plot size of two adjacent molecular clouds,   Rosette (1.46 pc) and Mon R2 (0.86  kpc), according to their masses \citep{1995ApJ...451..252W,2016MNRAS.461...22P}. The 5\% systematic error is included in the error bars. \label{fig:faceon}}
\end{figure*}

\section{Discussion}
\label{sec:discuss}

In this section, we compare our derived distances with previous results. Figure \ref{fig:faceon} displays a face-on view of the 11 molecular clouds, together with masers \citep{2018A&A...616L..15X} and local molecular clouds \citep{2019A&A...624A...6Y}. Masers are colored orange (the Local Arm), blue (the Perseus Arm), cyan (the Sagittarius Arm), green (the Outer Arm), and black (others). The sizes of the 11 molecular clouds (red circles) and two adjacent molecular clouds (purple), Rosette (1.46 kpc)  and Mon R2 (0.86 kpc), are proportional to their masses. The masses of Rosette \citep{1995ApJ...451..252W} and Mon R2 \citep{2016MNRAS.461...22P} molecular clouds are rescaled with updated distances, and are $1.3\times10^5$ and $4.3\times10^4$ \msun\ respectively.


\begin{figure*}[ht!]
\plotone{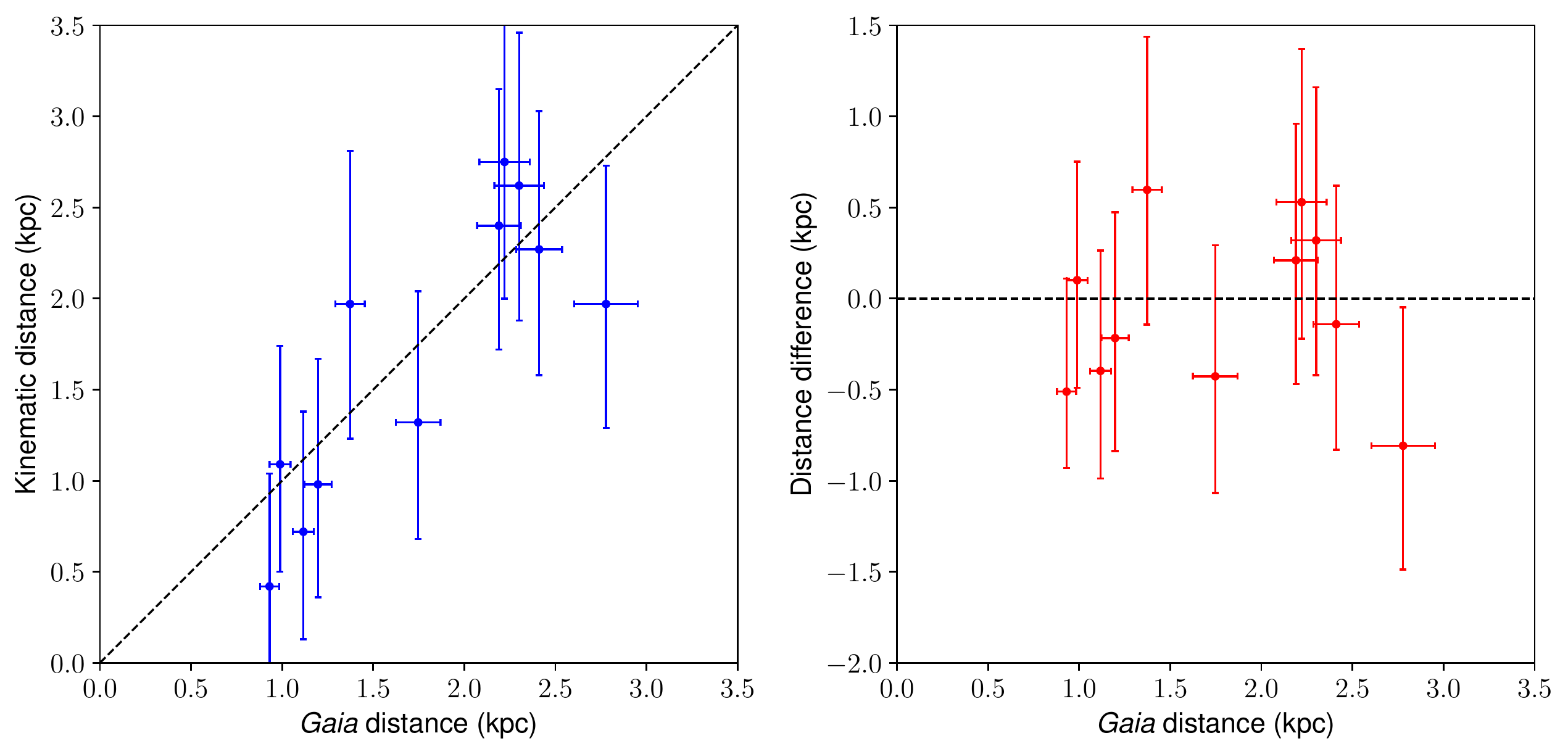}
\caption{A comparison of \textit{Gaia} and kinematic distances.  \label{fig:comparedis}}
\end{figure*}

\subsection{Individuals}

Five molecular clouds are located in the Perseus Arm ($D>2$ kpc), while the rest are in the Local Arm or the interarm region. In the Perseus Arm, the largest molecular cloud is G216 (G216.2-02.5), aka the Maddalena molecular cloud. The region of G216 is too large and, instead, we used two subregions, G216.2-02.5 and  G217.8-02.2. However, the \ag\ caused by G217.8-02.2 (the smaller one) is inadequate to reveal its distance, and we ignored this part and derived the mass and distance of G216 based on the region of G216.2-02.5. The total mass of G216 is about $3.1\times10^5$ \msun, consistent with the result of \citet{1994ApJ...432..167L}, while its distance  ($2411_{-72}^{+72}$ pc) is about 200 pc farther than previous results. \citet{1991ApJ...379..639L} derived a distance of 2.2 kpc for G216 using color excess of 42 stars. \citet{2019ApJ...879..125Z} derived a distance of 2.1 kpc, but they found that the ``ramp'' in foreground extinction makes the extinction distance underestimated, which is supported by our results. Around G216, four molecular clouds show similar distances, further confirming the distance of the Perseus Arm. For instance, the distance of S287 (G217.7-00.2) is 2.19 kpc, which agrees  with the study of \citet{1998ApJ...509..749H} (2.3 kpc).

However, two molecular clouds, G214.4-04.3 and G214.6-01.8, show large distance discrepancies. G214.4-04.3, which looks like a tail of G216, is the farthest molecular cloud of the 11 molecular clouds. Although its radial velocity is smaller than that of G216, G214.4-04.3 has a farther distance, 2.78 kpc, about 800 pc farther than its kinematic distance. This cloud may have a large systematic distance error,  due to its far distance and complicated foreground \ag\ variation. However, kinematic distances of the rest 10  molecular clouds agree with the \textit{Gaia} distances within errors. In addition to G214.4-04.3, G214.6-01.8, which harbors an molecular outflow (S287N), has a distance of 2.3 kpc, which is about 1 kpc farther than that given by \citet{1994ApJS...94..615H}. A comparison of \textit{Gaia} and kinematic distances is displayed in Figure \ref{fig:comparedis}.

The distance of the Outer Arm is too far to be determined. For instance, S284, at about 4.5 kpc \citep{2015A&A...584A..77N}, is certainly beyond the scope of \textit{Gaia} DR2. Indeed, no deviations were found between the \ag\ of on- and off-cloud stars for S284. Moving toward the fourth Galactic quadrant, the Outer Arm shows even far distances. For example, \citet{2008AJ....135..605S} determined the geometric distance (6.1 $\pm$ 0.6 kpc) for a  variable star,  V838 Monocerotis, which shows CO emission  at 53.3 \kms\ \citep{2008A&A...482..803K}. Clearly,  V838 Monocerotis is about 1.6 kpc farther than S284.

\subsection{The \ag\ baseline subtraction method}

The \ag\ baseline subtraction method makes many molecular cloud distances calculable. In the Galactic plane, it is almost impossible to derive correct distances to many molecular clouds using only on-cloud stars. We have shown that in many cases, off-cloud stars are able to trace the complicated  pattern of \ag\ contamination. Obviously, this baseline subtraction method also apply to extinction at other wavelengths, such as $A_V$ \citep{2019A&A...628A..94A}.
 

However, this method has limitations mainly due to the large uncertainties in \ag\ and parallaxes. The statistical analysis is only robust when sufficient numbers of both foreground and background on-cloud stars are present. Due to the extinction and parallax errors, background stars are usually much fewer than foreground stars for distant molecular clouds. If the number of background stars is too small, the MCMC process does not converge and the distance calculation is erroneous. We found that  molecular clouds farther than 3 kpc cannot be reliably analyzed with \textit{Gaia} DR2. 

In another situation, when the \ag\ of background stars is too high and is truncated, the \ag\ baseline subtraction method is also inapplicable. This usually happens in high-mass star forming regions, where the off-cloud \ag\ is likely to be high, and the baseline subtraction procedure would diminish the jump point of the on-cloud \ag. Consequently, the \ag\ baseline subtraction method cannot reliably handle molecular clouds with high column densities, in which case the truncated Gaussian method \citep{2019A&A...624A...6Y} should be used.

The  method assumes that the \ag\ of on- and off-cloud stars follows the same pattern if molecular clouds were not present. However, this assumption may not apply to regions where adjacent molecular clouds are unevenly distributed. It is even possible that the \ag\ of off-cloud stars exceed that of on-cloud stars because of the presence of other molecular clouds in the off-cloud region. In this case, the on- and off-cloud stars should be further selected to remove those stars that have been affected by foreground molecular clouds.

\subsection{Cloud boundaries and distance errors}

Accurately classifying on- and off-cloud stars is essential in the distance calculation. \citet{2019A&A...624A...6Y} used Planck 857 GHz continuum emission to trace molecular clouds, which is fine at high Galactic latitudes, but due to bright background emission, continuum emission is not qualified to do this task at low Galactic latitudes. 

In this work, we use footprints of molecular clouds in CO data cubes and CO integrated intensity maps to define on- and off-cloud regions. This is based on the assumption that molecular clouds have distinguishing velocities along the line of sight, which is not the case in velocity crowding regions, particularly in the Galactic anticenter direction. If two molecular clouds that have close Galactic coordinates and radial velocities (but different distances) were identified as one molecular cloud, we would see double jump positions in \ag\ toward the overlapping area. For other areas in the on-cloud region, we would see one single jump position caused by either of the molecular clouds. However, the  mixture of on-cloud stars that have different jump positions could cause a failure in distance calculation.

Although the cloud boundaries defined with CO is more accurate than with continuum emission, the  5\% systematic error is still present in our derived distances  due to  other issues, such as the large uncertainties in \ag, the choice of signal levels, insufficient number of backgrounds stars, misclassification of CO-dark molecular clouds, and the intrinsic structure of molecular clouds. For instance, the parallax was used to derive \ag\ \citep{2018A&A...616A...8A} and  the propagation of parallax systematic errors would cause systematic errors in \ag, thus affecting molecular cloud distances the second time. Although part of the systematic error in \ag\ would be removed in the baseline subtraction process, for far molecular clouds ($\sim$ 2.5 kpc), the distance error caused by the systematic parallax error \citep[0.03 mas,][]{2018A&A...616A...2L} is as large as 8\%, which could be further augmented by the systematic error of \ag. The systematic error is on the same order with statistical errors, and the next release of \textit{Gaia} data is expected to improve the accuracy of molecular cloud distances.

\subsection{Molecular clouds and spiral arms}

Remarkably, the location of the Perseus Arm follows the pattern traced by masers (blue) in the outer Galaxy, and the molecular cloud distances are also consistent with the dust distribution derived by \citet{2019arXiv190502734G}. There are a few molecular clouds in the interarm region. For instance, the Rosette molecular cloud (1.46 kpc) is located on the near edge of the Perseus Arm, and from Mon R2 (0.86 kpc) to the Rosette molecular cloud, the distribution of molecular clouds is  consecutive. This supports that the Milky Way may have no grand design pattern \citep{2016SciA....2E0878X}. 

Additional molecular cloud distances are needed to delineate the molecular cloud pattern in the Milky Way. In the 100 deg$^2$ region observed, the Perseus Arm has little foreground, so we can measure molecular clouds at about 2.5 kpc. However, toward regions where local molecular environments are complicated, we may  be unable to see as far as in the third Galactic quadrant. Nonetheless, \textit{Gaia} DR2 is able to determine molecular cloud distances in the Local and Perseus Arm up to approximately 2.5 kpc.

\section{Summary}
\label{sec:summary}

We have developed a new baseline subtraction method to calculate molecular cloud distances in the Galactic plane. We found that off-cloud \ag\ is suitable to calibrate on-cloud \ag\ and successfully determined distances to eleven molecular clouds in the third Galactic quadrant  using Bayesian analysis and MCMC sampling. Toward regions free of severe foreground contamination, \textit{Gaia} DR2 is capable of deriving distances for   molecular clouds at about  2.5 kpc.

We determined the location of the Perseus Arm, delineated by masers, using distances of molecular clouds, and confirmed the presence of molecular clouds in the interarm region. The completeness of the MWISP CO survey and future \textit{Gaia} data releases are expected to provide a fine picture for the distribution of local molecular clouds.

\acknowledgments

We thank Sam McSweeney for his careful proofreading. This work was sponsored by the Ministry of Science and Technology (MOST) Grant No. 2017YFA0402701, Key Research Program of Frontier Sciences (CAS) Grant No. QYZDJ-SSW-SLH047, the National Science Foundation of China (under Grand No. 11773077, 11873019, and 11673066), and China Postdoctoral Science Foundation under Grand No. 2018M642354.

%

\vspace{5mm}
\facilities{ PMO: 13.7m, \textit{Gaia}. }


\software{astropy \citep{2013A&A...558A..33A},  
          GILDAS \citep{2005sf2a.conf..721P}, 
          Duchamp \citep{2012MNRAS.421.3242W}, 
          Miriad \citep{1995ASPC...77..433S},
          sklearn \citep{scikit-learn},
          astrodendro.
          }

\appendix

\setcounter{figure}{0}
 
\renewcommand\thefigure{A.\arabic{figure}}
\renewcommand{\theHtable}{A.\thetable}

\newcommand{\capother}{The 5\% systematic error is not considered  and see the caption of Figure \ref{fig:g211dis} for other details.}

\section{Distances of four molecular clouds}
\label{}




 \begin{figure}[htb!]
 \figurenum{A.1}
\plotone{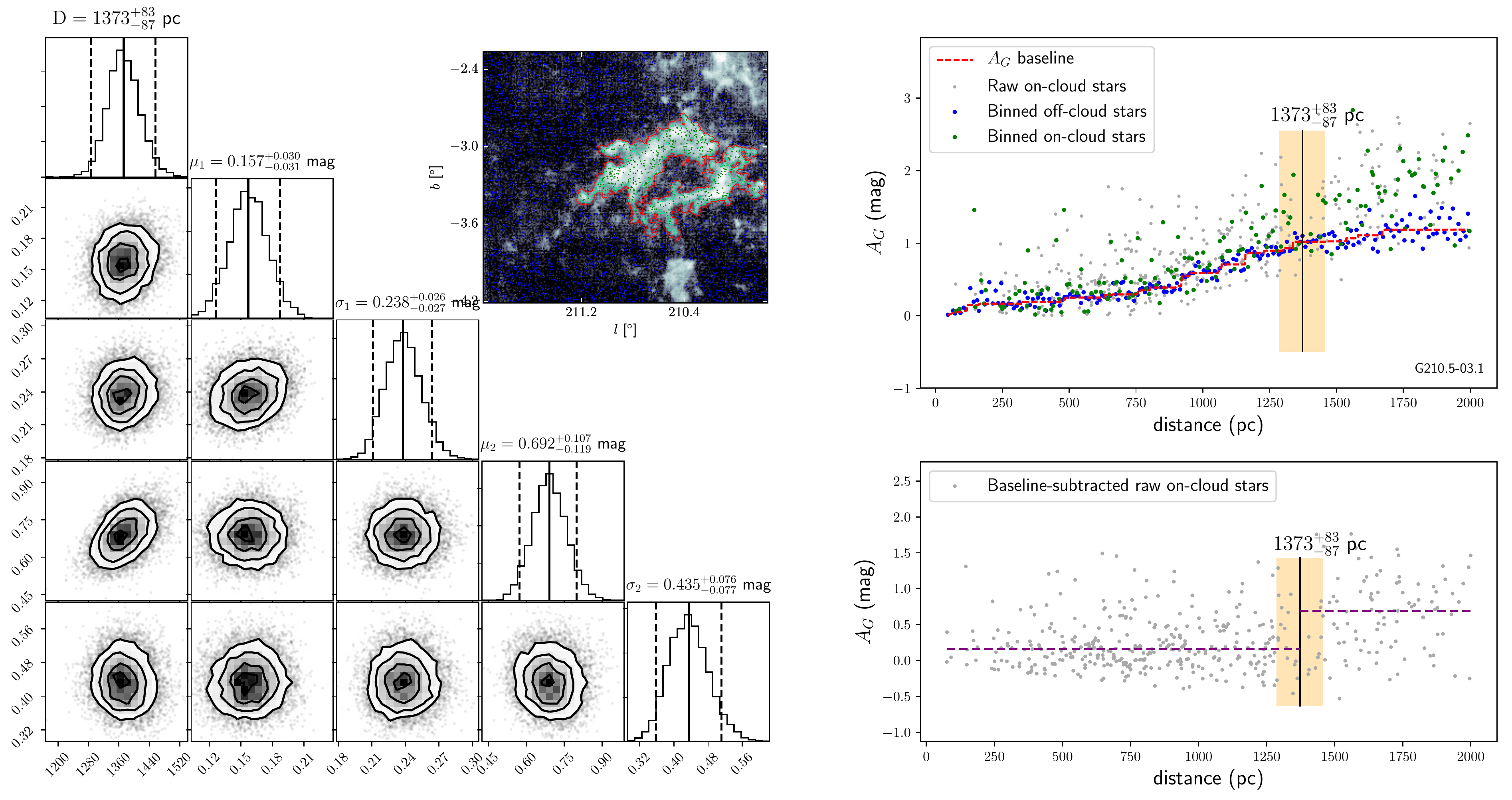}
\caption{ The distance of G210.5-03.1.   \capother   \label{fig:g2105dis}}
\end{figure}

\begin{figure}[htb!]
\figurenum{A.2}
\plotone{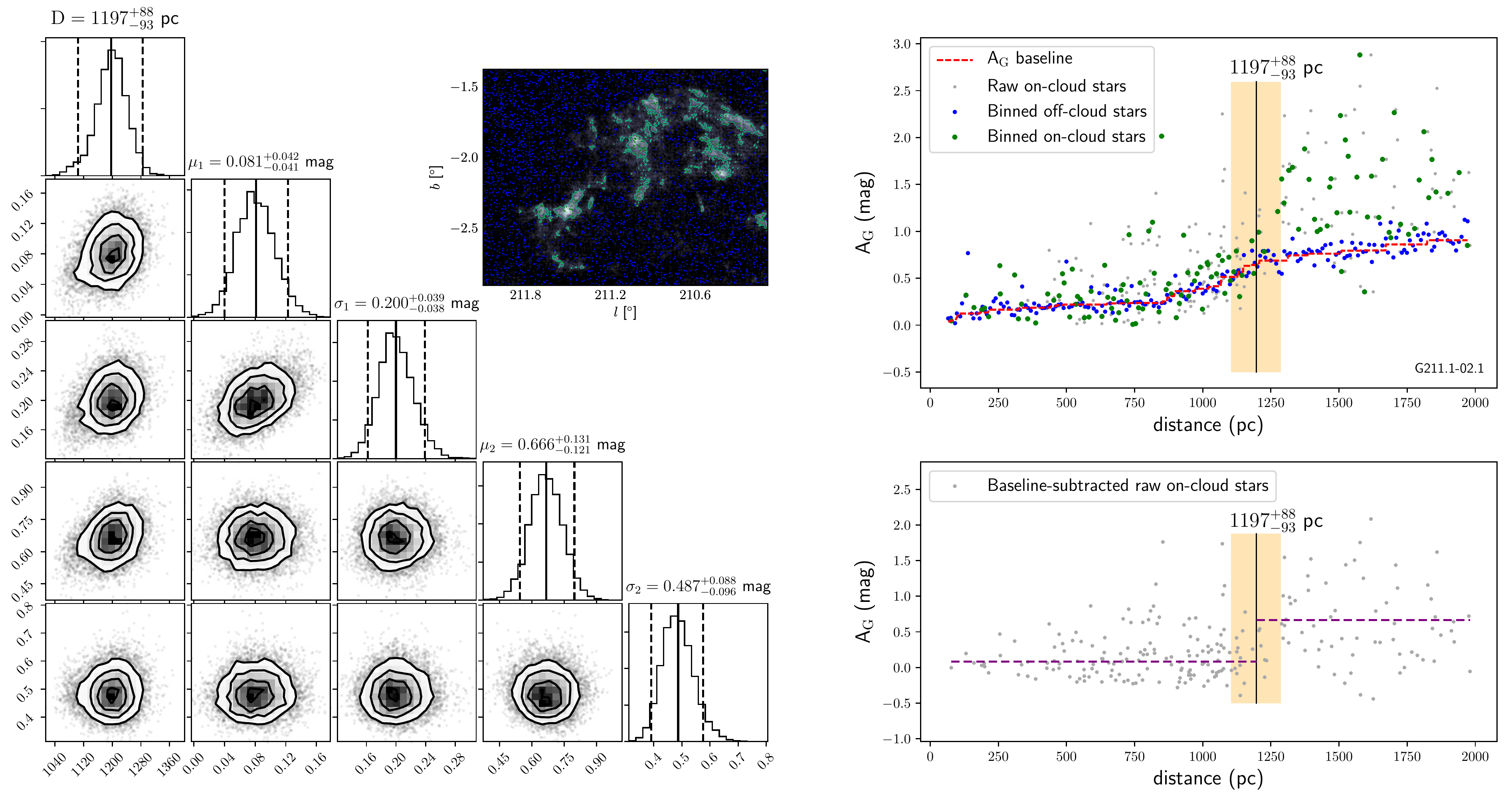}
\caption{The distance of G211.1-02.1. \capother  \label{fig:g2111dis}}
\end{figure}

 \begin{figure}[htb!]
 \figurenum{A.3}
\plotone{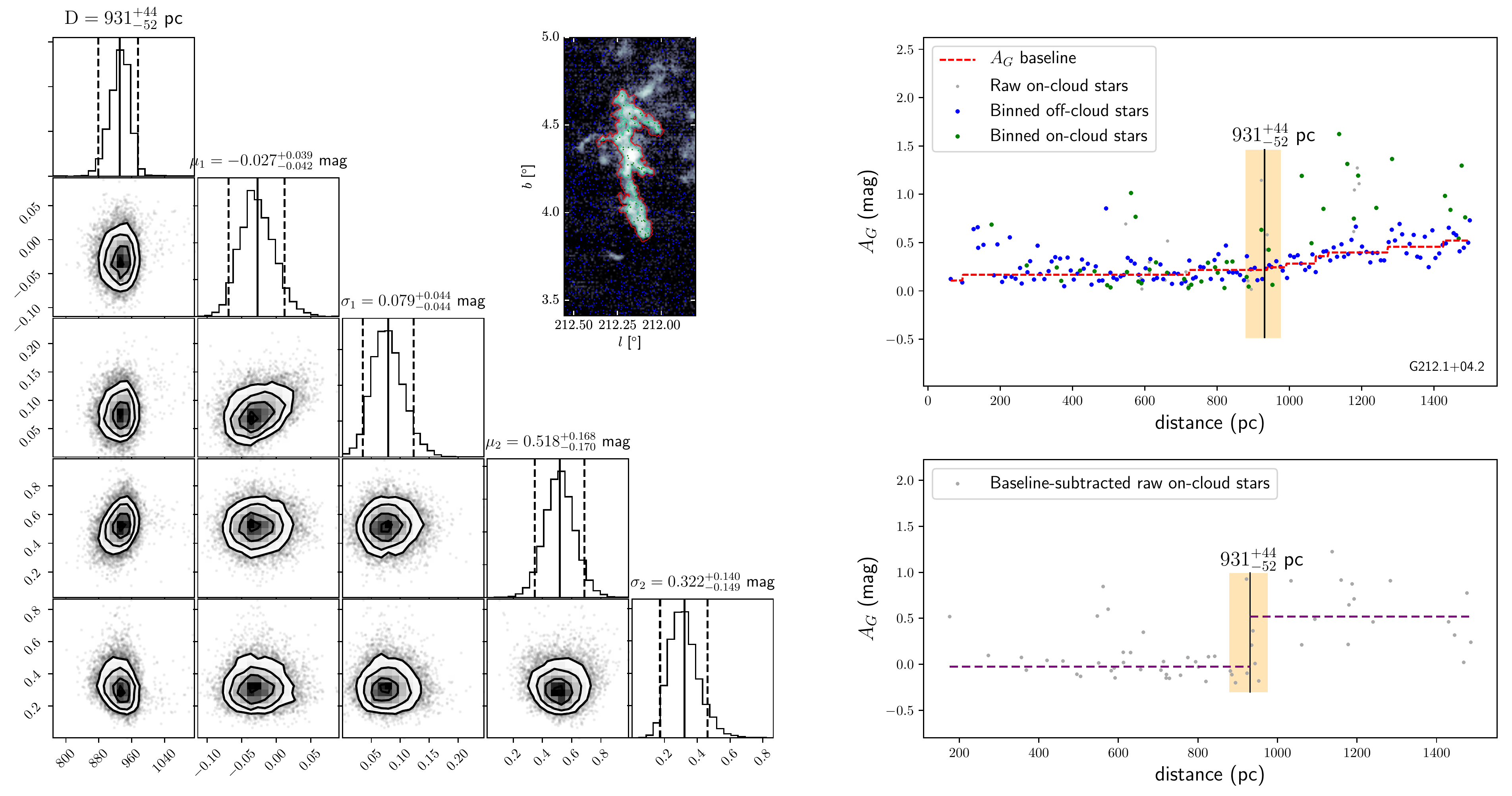}
\caption{ The distance of G212.1+04.2.  \capother \label{fig:G2121dis}}
\end{figure}

 \begin{figure}[htb!]
 \figurenum{A.4}
\plotone{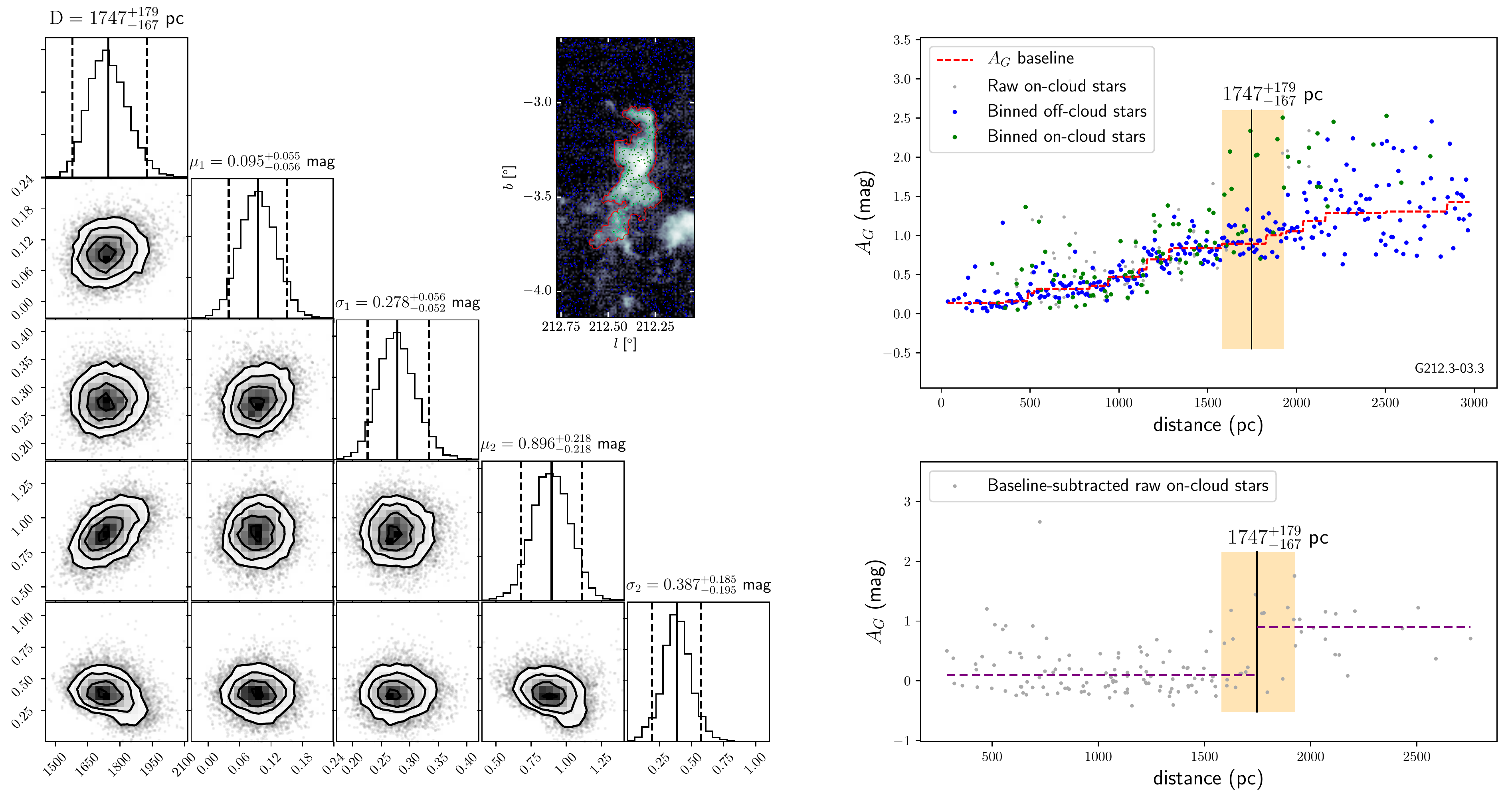}
\caption{ The distance of G212.3-03.3.   \capother   \label{fig:g2123dis}}
\end{figure}

 \begin{figure}[htb!]
 \figurenum{A.5}
\plotone{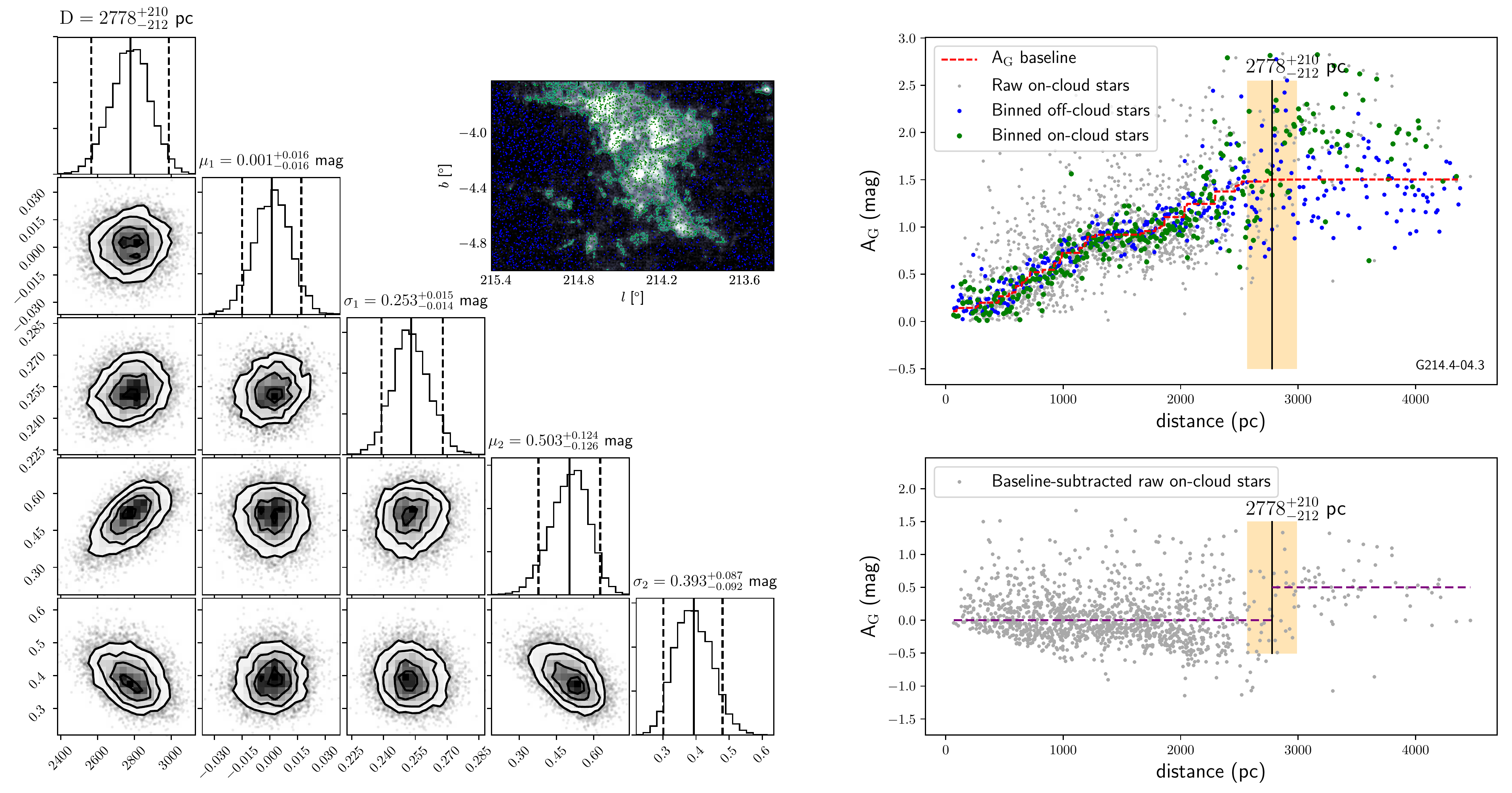}
\caption{ The distance of G214.4-04.3.   \capother  \label{fig:g2144dis}}
\end{figure}

 \begin{figure}[htb!]
 \figurenum{A.6}
\plotone{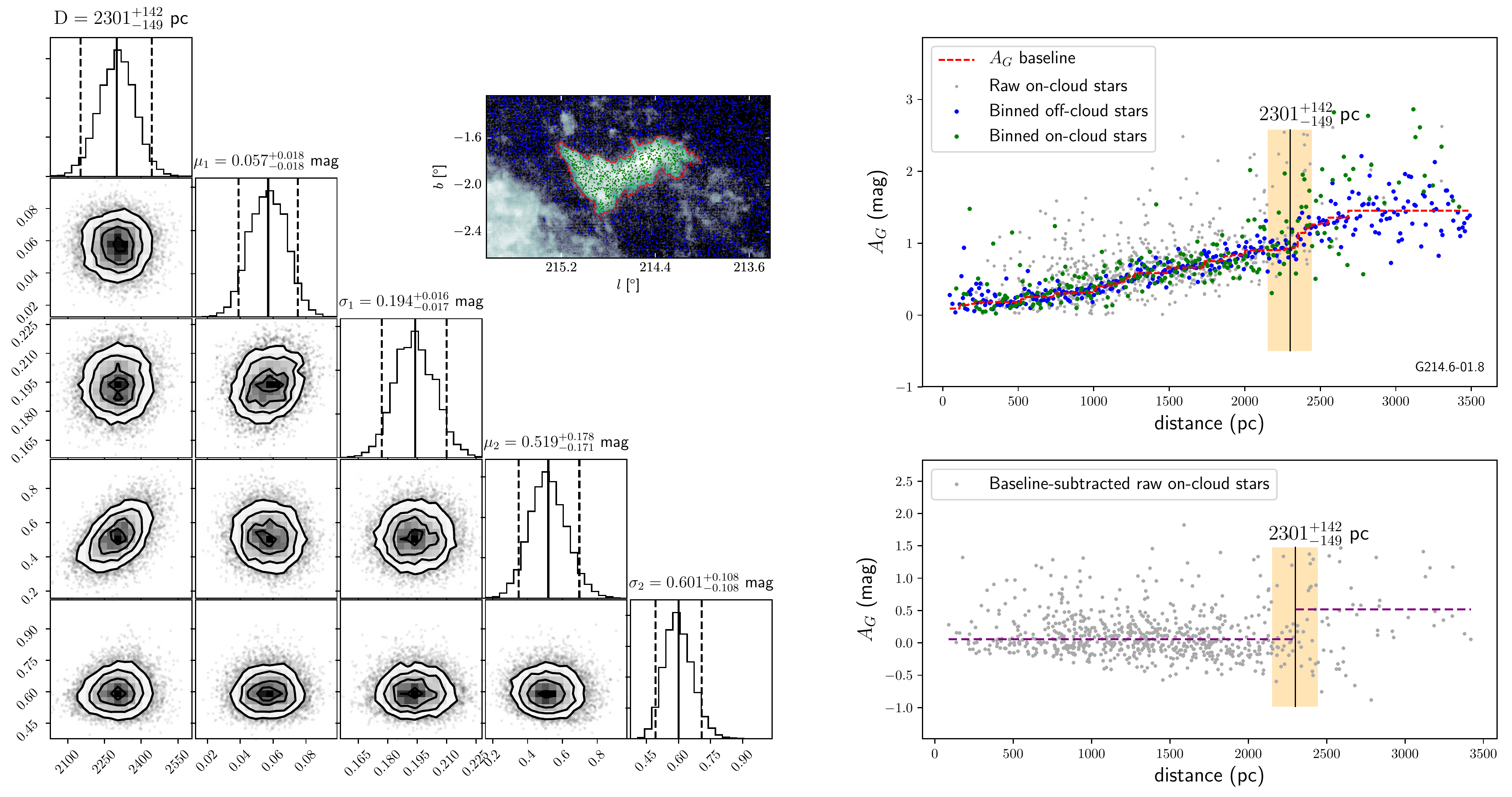}
\caption{ The distance of G214.6-01.8.   \capother \label{fig:g2146dis}}
\end{figure}

  \begin{figure}[htb!]
 \figurenum{A.7}
\plotone{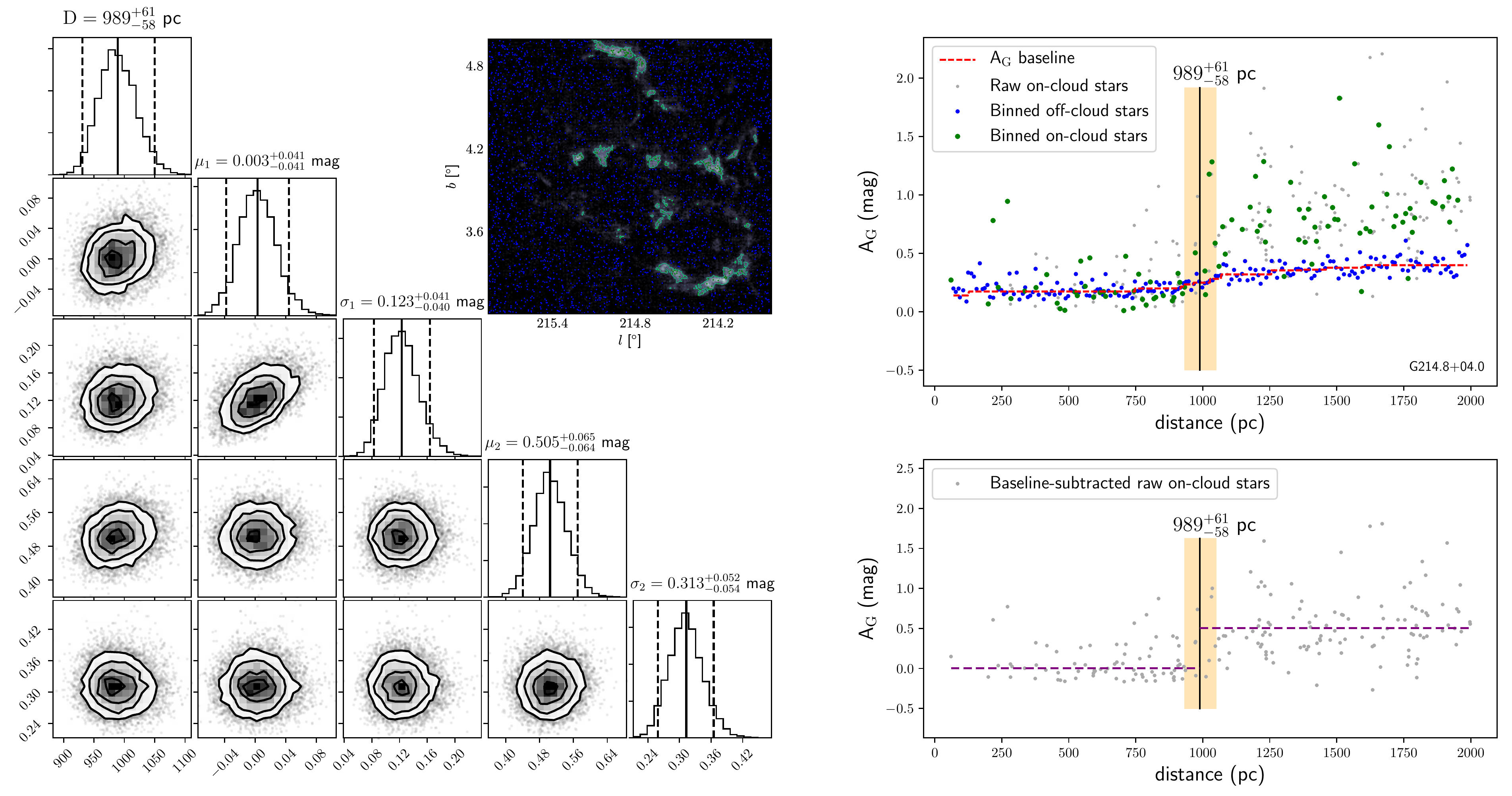}
\caption{ The distance of G214.8+04.0. \capother \label{fig:g2148dis}}
\end{figure}

 \begin{figure}[htb!]
 \figurenum{A.8}
\plotone{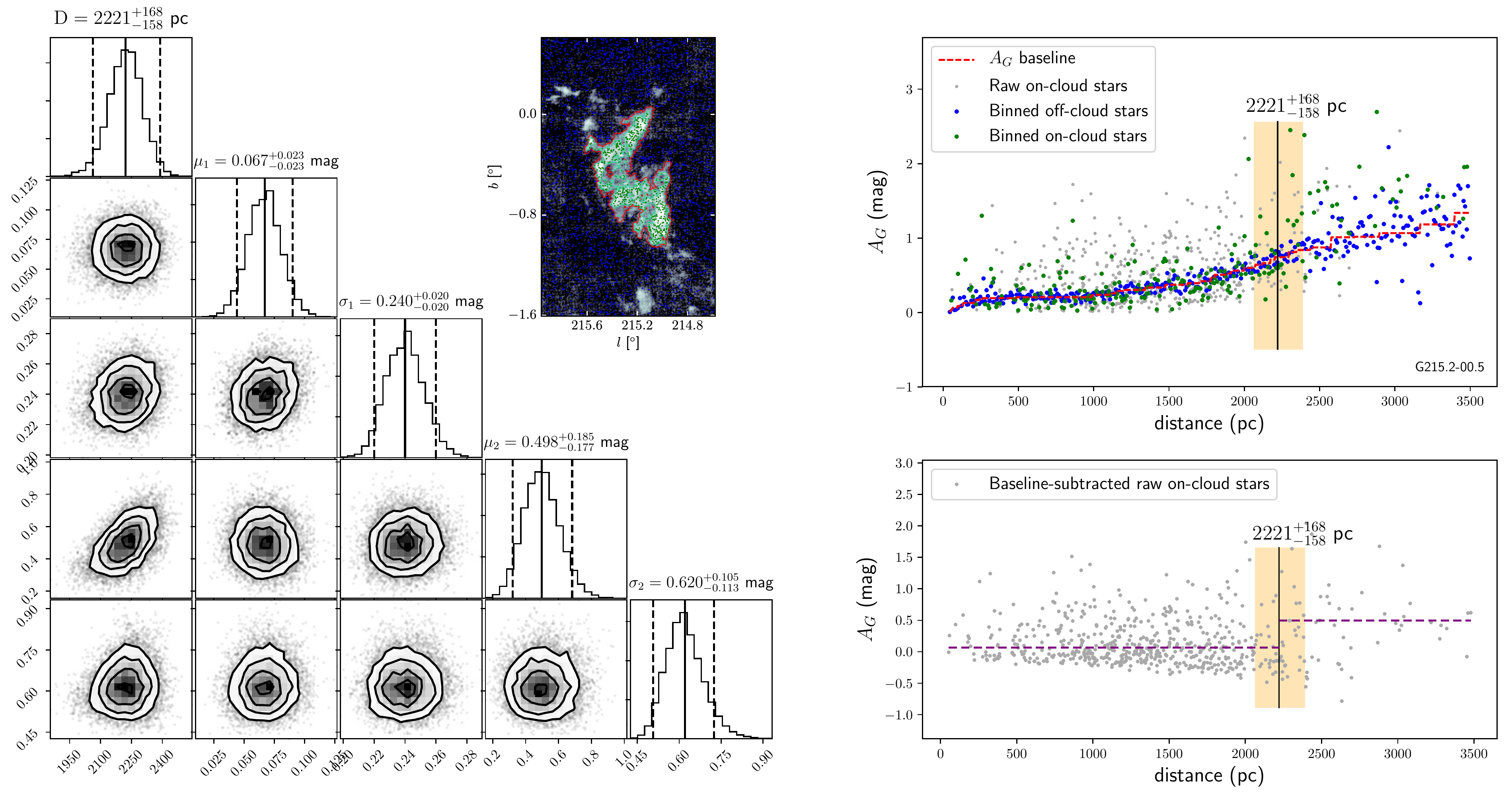}
\caption{ The distance of G215.2-00.5.   \capother   \label{fig:g2152dis}}
\end{figure}

 \begin{figure}[htb!]
 \figurenum{A.9}
\plotone{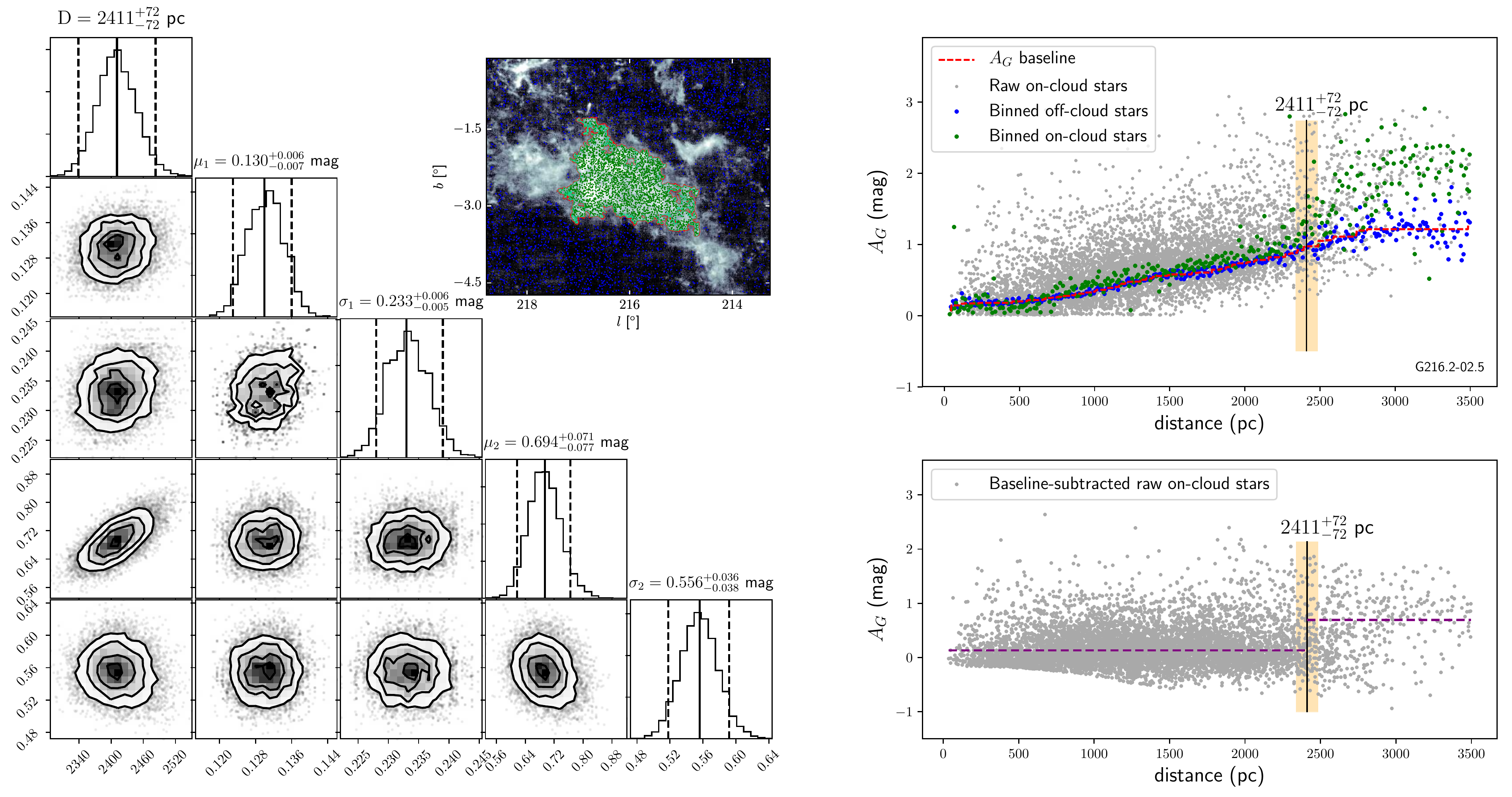}
\caption{ The distance of G216.2-02.5.  \capother \label{fig:g2162dis}}
\end{figure}

 \begin{figure}[htb!]
 \figurenum{A.10}
\plotone{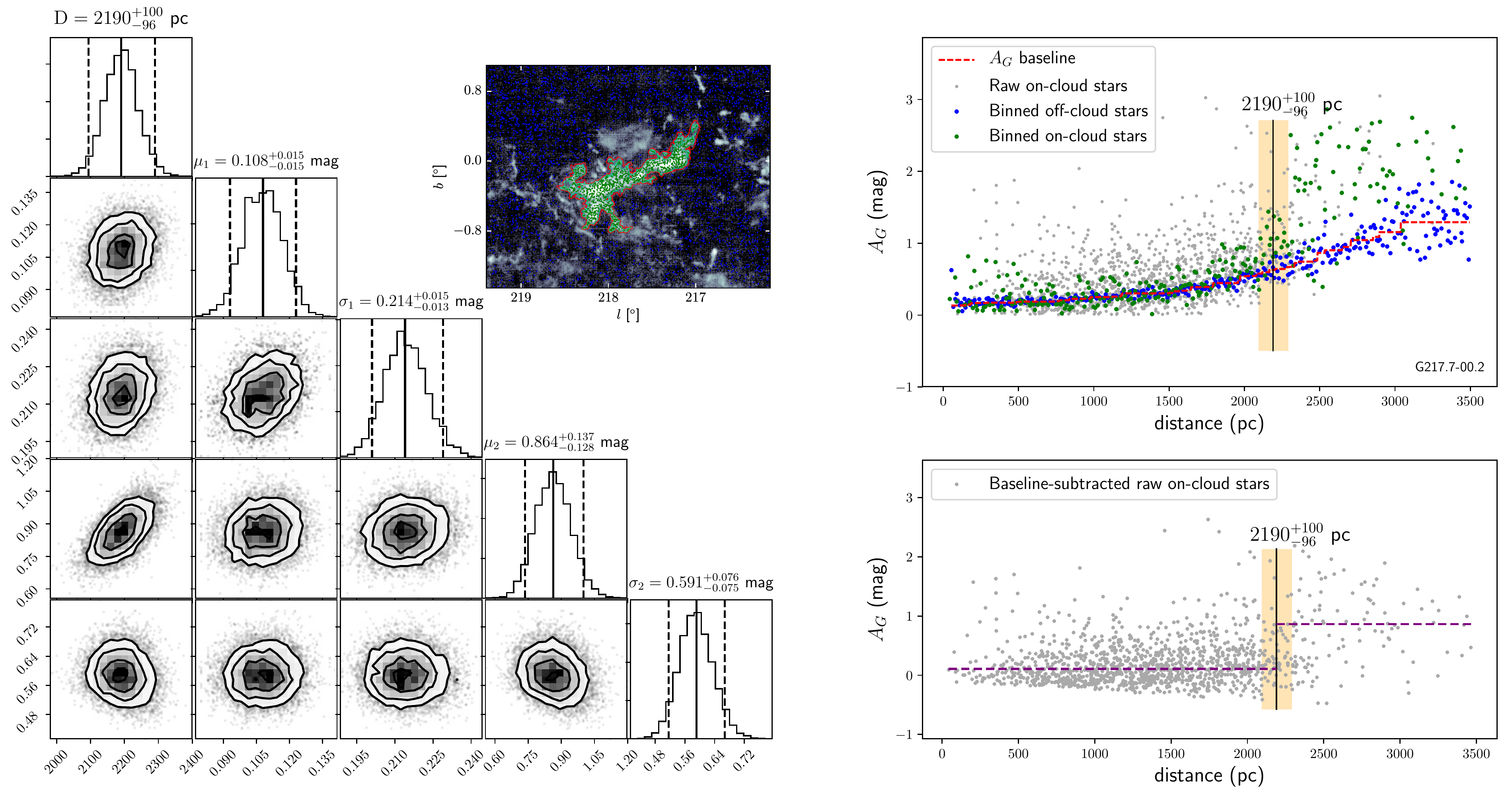}
\caption{ The distance of G217.7-00.2.  \capother  \label{fig:g2177dis}}
\end{figure}

\newpage



%

\end{document}